\documentstyle{mn2e}
\input psfig

  \makeatletter
  \ifx\CUP@mtlplain@loaded\undefined  
  \makeatother  
    
  \else
    
  \fi

\loadboldmathitalic 
\loadboldgreek      
  
\makeatletter
\ifx\CUP@mtlplain@loaded\undefined  
  \newfont\bit{cmbxti10 at 9pt}
\else
  \newfont\bit{mtbxti10 at 9pt}
\fi
\makeatother

\def\LaTeX{L\kern-.36em\raise.3ex\hbox{a}\kern-.15em
    T\kern-.1667em\lower.7ex\hbox{E}\kern-.125emX}

\newcommand{\gsim}{\mathrel{\hbox{\rlap{\lower.55ex \hbox {$\sim$}}
                   \kern-.3em \raise.4ex \hbox{$>$}}}}
\newcommand{\lsim}{\mathrel{\hbox{\rlap{\lower.55ex \hbox {$\sim$}}
                   \kern-.3em \raise.4ex \hbox{$<$}}}}
\newcommand{\msun}{\mbox{M$_\odot$}}

\newcommand{\mnras}{MNRAS}
\newcommand{\apj}{ApJ}

\newcommand{\aap}{A\&A}
\newcommand{\pasp}{PASP}
\newcommand{\pasj}{PASJ}
\newcommand{\aj}{AJ}

\title[The dependence of the IMF on the Jeans mass]{The origin of the initial mass function and its dependence on the mean Jeans mass in molecular clouds}

\author[M. R. Bate \& I. A. Bonnell]
  {Matthew R. Bate,$^1$\thanks{E-mail: mbate@astro.ex.ac.uk}
  Ian A. Bonnell.$^2$\\
  $^1$School of Physics, University of Exeter, Stocker Road,
    Exeter EX4 4QL \\
  $^2$School of Physics and Astronomy, University of St Andrews, North Haugh, St Andrews, Fife, KY16 9SS
}

\date{Accepted for publication in MNRAS}

\begin{document}
\maketitle

\begin{abstract}
We investigate the dependence of stellar properties on the mean thermal Jeans mass in molecular clouds.  We compare the results from the two largest hydrodynamical simulations of star formation to resolve the fragmentation process down to the opacity limit, the first of which was reported by Bate, Bonnell \& Bromm.  The initial conditions of the two calculations are identical except for the radii of the clouds, which are chosen so that the mean densities and mean thermal Jeans masses of the clouds differ by factors of nine and three, respectively.

We find that the denser cloud, with the lower mean thermal Jeans mass, produces a higher proportion of brown dwarfs and has a lower characteristic (median) mass of the stars and brown dwarfs.  This dependence of the initial mass function (IMF) on the density of the cloud may explain the observation that the Taurus star-forming region appears to be deficient in brown dwarfs when compared with the Orion Trapezium cluster.  The new calculation also produces wide binaries (separations $>20$ AU), one of which is a wide binary brown dwarf system.  

Based on the hydrodynamical calculations, we develop a simple accretion/ejection model for the origin of the IMF.  In the model, all stars and brown dwarfs begin with the same mass (set by the opacity limit for fragmentation) and grow in mass until their accretion is terminated stochastically by their ejection from the cloud through dynamically interactions.  The model predicts that the main variation of the IMF in different star-forming environments should be in the location of the peak (due to variations in the mean thermal Jeans mass of the cloud) and in the substellar regime.  However, the slope of the IMF at high-masses may depend on the dispersion in the accretion rates of protostars.
\end{abstract}


\section{Introduction}

Understanding the origin of the stellar initial mass function (IMF) is one of 
the fundamental goals of a complete theory of star formation.  
In the fifty years since Salpeter \shortcite{Salpeter1955} 
published his seminal paper on the form of the IMF, observational
studies have continually refined our knowledge of the IMF, 
largely overcoming the problems associated with the short lifetimes of 
high-mass stars, statistical significance, and stellar evolutionary models.  
Lately, observations have begun to determine the form of the IMF 
in the brown dwarf regime.  However, despite this progress in observationally
determining the form of the IMF, there is still no standard model for its 
origin, to say nothing of agreement on how it should depend on environment.

Many theories have been proposed for the origin of the IMF.  These fall into 
four main classes.  The IMF may originate from fragmentation, whether it be turbulent fragmentation (Larson 1992; Henriksen 1986, 1991; Elmegreen 1997, 1999, 2000b; Padoan, Nordlund \& Jones 1997; Padoan \& Nordlund 2002), gravitational fragmentation (Larson 1973; Elmegreen \& Mathieu 1983; Zinnecker 1984; Yoshii \& Saio 1985), or domain packing (Richtler 1994), with the fragmentation subject to an opacity limit which sets a minimum stellar mass (Hoyle 1953; Gaustad 1963; Yoneyama 1972; Suchkov \& Shchekinov 1976; Low \& Lynden-Bell 1976; Rees 1976; Silk 1977a, 1977b; Masunaga \& Inutsuka 1999).  It may depend on feedback processes (Shu et al.\ 1988; Silk 1995; Adams \& Fattuzzo 1996).  It may originate from competitive accretion of fragments (Hoyle 1953; Larson 1978; Zinnecker 1982; Bonnell et al. 1997, 2001a,b; Bonnell, Bate \& Vine 2003; Klessen, Burkert \& Bate 1998; Myers 2000). Or it may be due to coalesence or collisional build-up (Silk \& Takahashi 1979; Pumphrey \& Scalo 1983; Bonnell, Bate \& Zinnecker 1998; Bonnell \& Bate 2002).  In reality, all of these processes are likely to play some role.  The main questions to answer are which process, if any, dominates the origin the the IMF, and how does the IMF vary with environment?

One of the primary characteristics of the IMF is the characteristic mass.  One possibility is that the characteristic mass originates from the typical Jeans mass in the progenitor molecular cloud.  This may be the thermal Jeans mass (Larson 1992), a magnetic critical mass, or a turbulent Jeans mass (Silk 1995).  A Jeans mass origin for the characteristic stellar mass has been backed up by some hydrodynamical calculations of the fragmentation of clumpy and turbulent molecular clouds in which it was found that the mean mass of the protostars was similar to the mean initial Jeans mass in the cloud (Klessen, Burkert \& Bate 1998; Klessen \& Burkert 2000, 2001; Klessen 2001).  Another possibility is that the characteristic mass is due to the opacity limit for fragmentation, which sets a lower limit to the mass of a `star' and all other objects have masses greater than this minimum mass (Hoyle 1953).

In this paper, we report on the results from two large-scale hydrodynamical calculations of the collapse and fragmentation of turbulent molecular clouds.  The calculations resolve down to the opacity limit for fragmentation and, thus, capture the formation of all stars and brown dwarfs.  Many results from the first of these calculations have already been published (Bate, Bonnell \& Bromm 2002a, 2002b, 2003).  This calculation followed the fragmentation of a turbulent 50-\msun\ cloud with a mean initial thermal Jeans mass of 1 \msun.  It produced 50 stars and brown dwarfs with a mass function that was in good agreement with the observed IMF.  Here, we report the results of a second calculation, identical to the first, except that the radius of the cloud was reduced such that the mean thermal Jeans mass was reduced by a factor of three.  From these two calculations, we investigate the origin of the stellar initial mass function and its dependence on the Jeans mass in the clouds.

The paper is structured as follows.  Section 2 briefly describes the numerical method and the initial conditions for the calculations.  The results are discussed in Section 3.  In Section 4, we discuss the implications of the results for the origin of the IMF.  Our conclusions are given in Section 5.

\section{Computational method}

The calculations presented here were performed using a three-dimensional, 
smoothed particle hydrodynamics (SPH) code.  The SPH code is 
based on a version originally developed by Benz 
(Benz 1990; Benz et al.\ 1990).
The smoothing lengths of particles are variable in 
time and space, subject to the constraint that the number 
of neighbours for each particle must remain approximately 
constant at $N_{\rm neigh}=50$.  The SPH equations are 
integrated using a second-order Runge-Kutta-Fehlberg 
integrator with individual time steps for each particle 
\cite{BatBonPri1995}.
Gravitational forces between particles and a particle's 
nearest neighbours are calculated using a binary tree.  
We use the standard form of artificial viscosity 
(Monaghan \& Gingold 1983; Monaghan 1992) with strength 
parameters $\alpha_{\rm_v}=1$ and $\beta_{\rm v}=2$.
Further details can be found in Bate et al.\ \shortcite{BatBonPri1995}.
The code has been parallelised by M.\ Bate using OpenMP.  

\subsection{Equation of state}
\label{eossec}

To model the thermal behaviour of the gas without performing radiative transfer,
we use a barotropic equation of state for the thermal pressure of the
gas $p = K \rho^{\eta}$, where $K$ is a measure of the entropy
of the gas.  The value of the effective polytropic exponent $\eta$, 
varies with density as
\begin{equation}\label{eta}
\eta = \cases{\begin{array}{rl}
1, & \rho \leq 10^{-13}~ {\rm g~cm}^{-3}, \cr
7/5, & \rho > 10^{-13}~ {\rm g~cm}^{-3}. \cr
\end{array}}
\end{equation}
We take the mean molecular weight of the gas to be $\mu = 2.46$.
The value of $K$ is defined such that when the gas is 
isothermal $K=c_{\rm s}^2$, with the sound speed
$c_{\rm s} = 1.84 \times 10^4$ cm s$^{-1}$ at 10 K,
and the pressure is continuous when the value of $\eta$ changes.
This equation of state has been chosen to match closely the 
relationship between temperature and density during the 
spherically-symmetric collapse of molecular 
cloud cores as calculated with frequency-dependent radiative 
transfer (see Bate et al.\ 2003 for further details).

\subsection{Sink particles}
\label{sinkparticles}

The heating of the molecular gas that begins at a density of $10^{-13}$ g~cm$^{-3}$ inhibits fragmentation at higher densities.  This is how we model the opacity limit for fragmentation.  The opacity limit for fragmentation results in the formation 
of distinct pressure-supported
fragments in the calculation.  As these fragments accrete, their
central density increases, and it becomes computationally impractical
to follow their internal evolution because of the short dynamical
time-scales involved.  Therefore, when the central density of 
a pressure-supported fragment exceeds 
$\rho_{\rm s} = 10^{-11}~{\rm g~cm}^{-3}$, 
we insert a sink particle into the calculation (Bate et al.\ 1995).

In the calculations presented here, a sink particle is formed by 
replacing the SPH gas particles contained within $r_{\rm acc}=5$ AU 
of the densest gas particle in a pressure-supported fragment 
by a point mass with the same mass and momentum.  Any gas that 
later falls within this radius is accreted by the point mass 
if it is bound and its specific angular momentum is less than 
that required to form a circular orbit at radius $r_{\rm acc}$ 
from the sink particle.  Thus, gaseous discs around sink 
particles can only be resolved if they have radii $\gsim 10$ AU.
Sink particles interact with the gas only via gravity and accretion.

\begin{table*}
\begin{tabular}{lccccccccc}\hline
Calculation & Initial Gas & Initial  & Jeans & Mach & No. Stars & No. Brown  & Mass of Stars and  & Mean & Median \\
 & Mass  & Radius & Mass & Number & Formed & Dwarfs Formed & Brown Dwarfs & Mass & Mass \\
 & M$_\odot$ & pc & M$_\odot$ & & & & M$_\odot$ & M$_\odot$ & M$_\odot$\\ \hline
1 & 50.0 & 0.188 & 1 & 6.4 & $\geq$23 & $\leq$27 & 5.89 & 0.1178 & 0.070 \\
2 & 50.0 & 0.090 & 1/3 & 9.2 & $\geq$19 & $\leq$60 & 7.92 & 0.1003 & 0.023 \\ \hline
\end{tabular}
\caption{\label{table1} The initial conditions for the two calculations and the statistical properties of the stars and brown dwarfs formed.  The initial conditions were identical except that Calculation 2 had a cloud with a smaller radius giving a mean thermal Jeans mass a factor of 3 lower.  Their initial turbulent velocity fields were identical except for their magnitudes which were scaled so that for both clouds the kinetic energy equalled the magnitude of the gravitational potential energy.  Both calculations were run for 1.40 initial cloud free-fall times.  Brown dwarfs are defined as having final masses less than 0.075 M$_\odot$.  The numbers of stars (brown dwarfs) are lower (upper) limits because some of the brown dwarfs were still accreting when the calculations were stopped.  The mean mass of the objects formed in Calculation 2 was 17\% lower than in Calculation 1 and the median mass was a factor of 3.04 lower (consistent with the change in the initial mean thermal Jeans mass).}
\end{table*}

Since all sink particles are created from pressure-supported 
fragments, their initial masses are a few Jupiter masses (M$_{\rm J}$), 
as given by the opacity limit for fragmentation 
(Low \& Lynden-Bell 1976; Rees 1976; Silk 1977a, 1977b).  
Subsequently, they may accrete large amounts of material 
to become higher-mass brown dwarfs ($\lsim 75$ M$_{\rm J}$) or 
stars ($\gsim 75$ M$_{\rm J}$), but {\it all} the stars and brown
dwarfs begin as these low-mass pressure-supported fragments.

The gravitational acceleration between two sink particles is
Newtonian for $r\geq 4$ AU, but is softened within this radius
using spline softening \cite{Benz1990}.  The maximum acceleration 
occurs at a distance of $\approx 1$ AU; therefore, this is the
minimum separation that a binary can have even if, in reality,
the binary's orbit would have been hardened.  Sink particles are
not permitted to merge in this calculation.

The benefits and potential problems associated with introducing sink particles are discussed in more detail by Bate et al.\ \shortcite{BatBonBro2003}.

\subsection{Initial conditions}

We report on the results from two calculations.  The initial conditions for the calculations are identical except for radii of the initial clouds.  They are summarised in Table \ref{table1}.  For each calculation, the initial conditions consist of a large-scale, turbulent molecular cloud.  Each cloud is spherical and uniform in density with a mass of 50 M$_\odot$.  For Calculation 1, the diameter of the cloud is $0.375$ pc (77400 AU), while for Calculation 2, the diameter is $0.180$ pc (37200 AU).  At temperatures of 10 K, the mean thermal Jeans masses are 1 M$_\odot$ in Calculation 1
(i.e.\ the cloud contains 50 thermal Jeans masses) and 1/3 M$_\odot$ in Calculation 2
(i.e.\ the cloud contains 150 thermal Jeans masses).
The free-fall times of the clouds are $t_{\rm ff}=6.0\times 10^{12}$~s
or $1.90\times 10^5$ years and $t_{\rm ff}=2.0\times 10^{12}$~s
or $6.34\times 10^4$ years, respectively.

Although the clouds are uniform in density, we impose an initial 
supersonic `turbulent' velocity field on them in the same manner
as Ostriker, Stone \& Gammie \shortcite{OstStoGam2001}.  We generate a
divergence-free random Gaussian velocity field with a power spectrum 
$P(k) \propto k^{-4}$, where $k$ is the wavenumber.  
In three dimensions, this results in a
velocity dispersion that varies with distance, $\lambda$, 
as $\sigma(\lambda) \propto \lambda^{1/2}$ in agreement with the 
observed Larson scaling relations for molecular clouds 
\cite{Larson1981}.
This power spectrum is slightly steeper than the Kolmogorov
spectrum, $P(k)\propto k^{-11/3}$.  Rather, it matches the 
amplitude scaling of Burgers supersonic turbulence associated
with an ensemble of shocks (but differs from Burgers turbulence
in that the initial phases are uncorrelated).
The velocity field is generated on a $64^3$ uniform grid and the
velocities of the particles are interpolated from the grid.  The {\it same velocity field is used for both calculations}, but the velocity field is normalised so that the kinetic energy of the turbulence equals the magnitude of the gravitational potential energy of each cloud.
Thus, the initial root-mean-square (rms) Mach number of the turbulence 
is ${\cal M}=6.4$ in Calculation 1 and ${\cal M}=9.2$ in Calculation 2.  
In some ways it is undesirable that the turbulent Mach number is different in the two calculations; it might have been preferable to vary only the thermal Jeans mass.  Maintaining the same Mach number while reducing the thermal Jeans mass could have been accomplished by reducing the radius {\it and mass} of the cloud used for Calculation 1 by the same fraction.  However, this would have reduced the total number of objects formed in Calculation 2.  Since neither calculation forms a very large number of objects and looking for statistically significant differences between the two calculations is difficult enough, it is likely that using a lower mass cloud for Calculation 2 would have resulted in any differences between the two calculations being statistically insignificant.  There is also an advantage to varying both the turbulent Mach number and the mean thermal Jeans mass.  That is, if a difference between the characteristic stellar masses of the two calculations is found, we can look to see whether the characteristic mass scales more closely with the mean thermal Jeans mass or the turbulent Jeans mass (see Section 3.5).

Note that the initial conditions for Calculation 2 are extreme both in terms of mean density and initial Mach number.  Such initial conditions are not found in nearby star-forming regions.  In order to study the dependence of star formation on the mean thermal Jeans mass in molecular clouds, it might have been more desirable to vary the cloud parameters in the opposite direction (i.e. increase the mean thermal Jeans mass by beginning with lower density, less turbulent clouds).  However, currently we are limited by the available computational time to studying clouds containing only 50 M$_\odot$ of gas (while resolving the opacity limit) and increasing the mean thermal Jeans mass would mean that we would no longer be in the regime of studying clouds containing many Jeans masses (i.e. likely to form large numbers of objects so that we can examine their statistical properties).  Thus, the primary purpose of Calculation 2 is to test the dependence of star formation on the mean thermal Jeans mass; the vast majority of star formation probably does not occur in such extreme environments.

\begin{table*}
\begin{tabular}{lccccccc}\hline
Core & Initial Gas  & Initial  & Final  & No. Stars & No. Brown  & Mass of Stars and  & Star Formation \\
 & Mass  & Size & Gas Mass & Formed & Dwarfs Formed & Brown Dwarfs & Efficiency  \\
 & M$_\odot$ & pc & M$_\odot$ &  & & M$_\odot$ & \% \\ \hline
1 & 1.50 (0.15) & $0.04\times 0.04\times 0.03$ & 2.03 (1.04) & $\geq$13 & $\leq$52 & 6.33 & 76 (86) \\
2 & 0.92 (0.16) & ($0.03\times 0.01\times 0.01$) & 1.18 (0.50) & $\geq$4 & $\leq$8 & 1.33 & 53 (73)\\
3 & 0.17 (0.06) & ($0.02\times 0.01\times 0.01$) & 0.32 (0.08) & 1 & 0 & 0.18 & 36 (69)\\
4 & 0.31 (0.07) & ($0.03\times 0.01\times 0.01$) & 0.32 (0.06) & 1 & 0 & 0.09 & 22 (60)\\\hline
Cloud & 50.0 & $0.38\times 0.38\times 0.38$ & 42.1 & $\geq$19 & $\leq$60 & 7.92 & 16 \\ \hline
\end{tabular}
\caption{\label{table2} The properties of the four dense cores that form during Calculation 2 and those of the cloud as a whole.  The gas masses and sizes of the cores are calculated from gas with $n({\rm H}_2)>1\times 10^7$~cm$^{-3}$ and $n({\rm H}_2)>1\times 10^8$~cm$^{-3}$ (the latter values are given in parentheses).  These densities are an order of magnitude higher than those used for Calculation 1 in BBB2003 because the cloud is nearly an order of magnitude denser.  The initial gas mass is calculated just before star formation begins in that core (i.e.\ different times for each core).  Brown dwarfs have final masses less than 0.075 M$_\odot$.  The star formation efficiency is taken to be the total mass of the stars and brown dwarfs that formed in a core divided by the sum of this mass and the mass in gas in that core at the end of the calculation.  As with Calculation 1, the star formation efficiency is high locally, but low globally.  The numbers of stars (brown dwarfs) are lower (upper) limits because fourteen of the brown dwarfs were still accreting when the calculation was stopped. }
\end{table*}

\subsection{Resolution}

The local Jeans mass must be resolved throughout 
the calculations to model fragmentation correctly 
(Bate \& Burkert 1997; Truelove et al.\ 1997; Whitworth 1998; 
Boss et al.\ 2000).  This requires $\gsim 1.5 N_{\rm neigh}$ SPH particles per Jeans mass; 
$N_{\rm neigh}$ is insufficient (Bate et al.\ 2003).
The minimum Jeans mass in the calculation presented here occurs 
at the maximum density during the isothermal phase of the 
collapse, $\rho = 10^{-13}$ g~cm$^{-3}$, 
and is $\approx 0.0011$ M$_\odot$ (1.1 M$_{\rm J}$).  Thus, we
use $3.5 \times 10^6$ particles to model the 50-M$_\odot$ clouds.
The calculations required approximately 95000 and 50000 CPU hours, respectively, on the SGI Origin 3800 of the United Kingdom Astrophysical Fluids Facility (UKAFF).

\section{Comparison of results}
\label{results}

The results of Calculation 1 were published in detail by Bate et al.\ (2002a,b; 2003), who considered the global evolution of the cloud, the star formation efficiency and timescale, the form of the stellar initial mass function, the formation mechanisms of brown dwarfs and close binaries, the multiplicity and velocity dispersion of the objects, and the properties of their circumstellar discs.  In this paper, we compare the results of the two calculations, under the same headings, to determine how the star formation process and the properties of stars and brown dwarfs depend on the mean Jeans mass of the progenitor molecular cloud.  In the text we concentrate on how the results {\it differ} between the two calculations, although the figures and tables in this paper provide the detailed results of Calculation 2 in an identical manner to those presented for Calculation 1 in Bate et al.\ \shortcite{BatBonBro2003}, hence forth referred to as BBB2003.

\begin{figure*}
\centerline{\psfig{figure=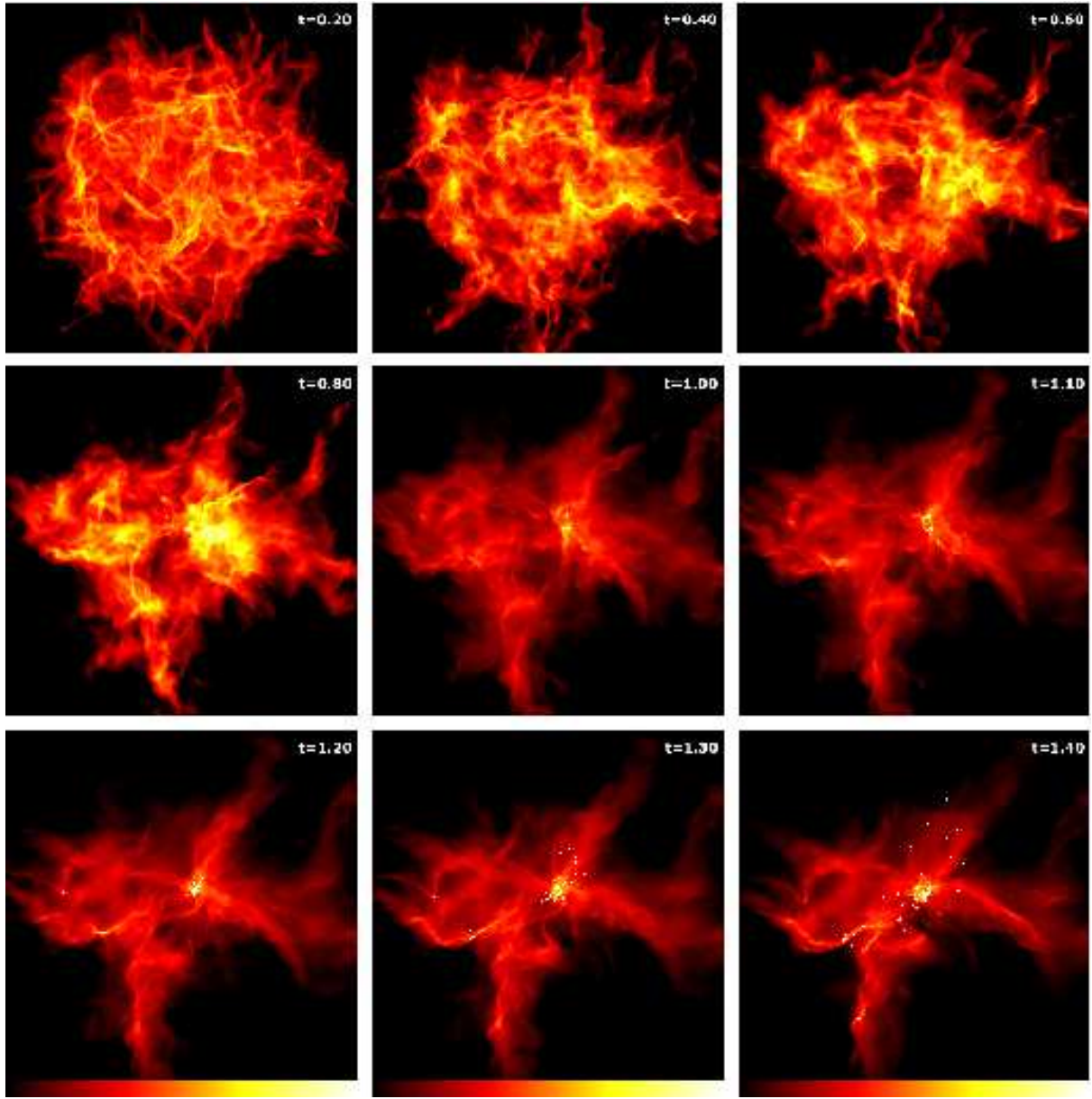,width=16.0truecm}}
\caption{\label{global} The global evolution of the cloud during Calculation 2 for comparison with Figure 2 of BBB2003 for Calculation 1.  Shocks lead to the dissipation of the turbulent energy that initially supports the cloud, allowing parts of the cloud to collapse.  Star formation begins at $t=0.82t_{\rm ff}$ in a collapsing dense core.  By the end of the calculation, three more dense cores have begun forming stars (left-hand side of the last panel) and many of the stars and brown dwarfs have been ejected from the cloud through dynamical interactions.  Each panel is 0.194 pc (40 000 AU) across.  Time is given in units of the initial free-fall time of $6.34\times 10^4$ yr.  The panels show the logarithm of column density, $N$, through the cloud, with the scale covering $-0.9<\log N<0.6$ for $t<1.0$ and $-1.1<\log N<2.1$ for $t\ge 1.0$ with $N$ measured in g~cm$^{-2}$.  This column density scale is chosen to allow direct comparison with Calculation 1.} 
\end{figure*}

\begin{figure*}
\centerline{\psfig{figure=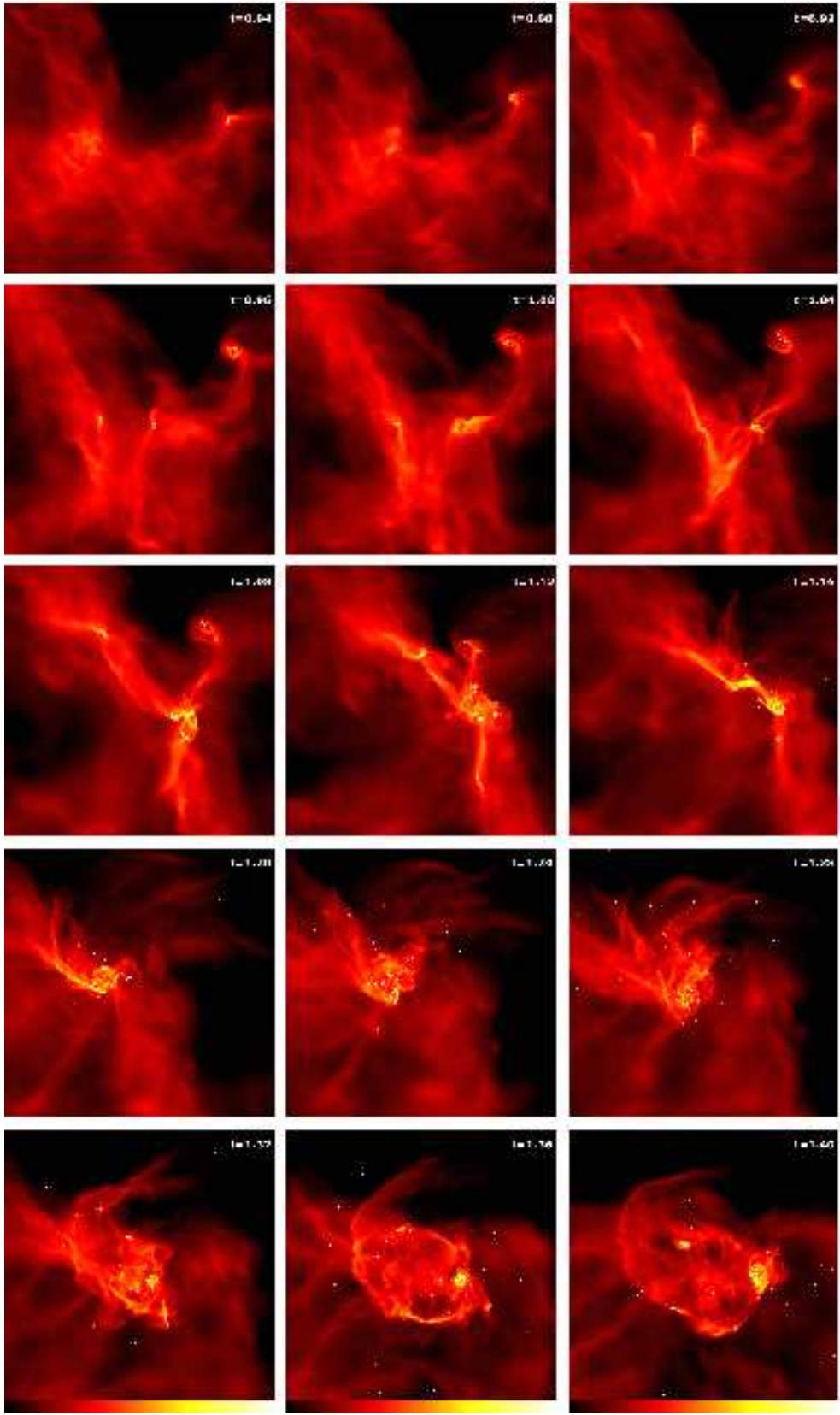,width=13.0truecm}}
\caption{\label{core1} The star formation in the first (main) dense core.  The first objects form a binary at $t=0.824t_{\rm ff}$.  Large gaseous filaments collapse to form single objects and multiple systems.  These objects fall together to form a small group.  The group quickly dissolves due to dynamical interactions, leaving behind a bound remnant.  Compared with Figures 3 and 4 of BBB2003 for Calculation 1, the stellar group formed here is denser and more numerous leading to more violent dynamical interactions and fewer large discs.  Each panel is 0.025 pc (5150 AU) across.  Time is given in units of the initial free-fall time of $6.34\times 10^4$ yr.  The panels show the logarithm of column density, $N$, through the cloud, with the scale covering $-0.2<\log N<2.5$ with $N$ measured in g~cm$^{-2}$. } 
\end{figure*}

\begin{figure*}
\centerline{\psfig{figure=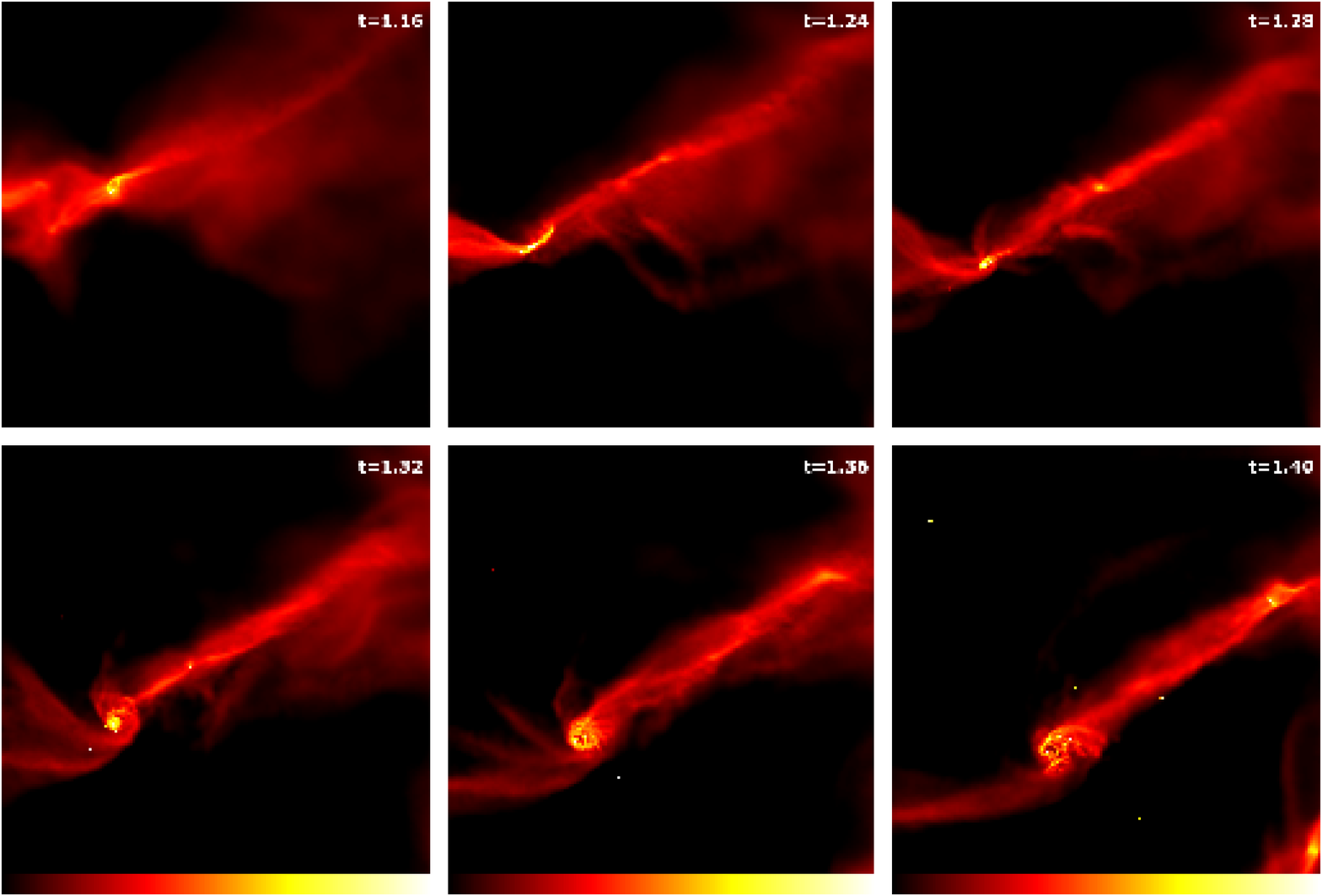,width=16.0truecm}}
\caption{\label{core2} The star formation in the second dense core.  The first object forms at $t=1.111t_{\rm ff}$, followed quickly by a second to form a binary which is then surrounded by a circumbinary disc.  The massive disc eventually undergoes fragmentation to form six more objects.  During this time, four more objects are also formed in filaments.  By the end of the calculation, three of the objects have been ejected from the core via dynamical interactions, leaving a hierarchical quadruple system surrounded by five brown dwarfs in wide unstable orbits.  Each panel is 0.025 pc (5150 AU) across.  Time is given in units of the initial free-fall time of $6.34\times 10^4$ yr.  The panels show the logarithm of column density, $N$, through the cloud, with the scale covering $-0.2<\log N<2.5$ with $N$ measured in g~cm$^{-2}$.} 
\end{figure*}

\subsection{Evolution of the clouds}
\label{evolution}

The global evolution of Calculation 2 is very similar to that reported for Calculation 1.  Although the cloud is initially uniform in density and the turbulent velocity field is initially divergence-free, hydrodynamic evolution quickly results in the formation of shocks, first on small scales, and later on larger scales.  Kinetic energy is lost from the cloud in these shocks, reducing the turbulent support, and gravity soon begins to dominate in regions of overdensity formed by converging gas flows.  As found in other calculations of turbulent molecular clouds, both with and without magnetic fields and self-gravity, the turbulence decays on the dynamical time-scale of the cloud (e.g. Mac Low et al.\ 1998; Stone, Ostriker \& Gammie 1998).  Thus, star formation begins after approximately one global free-fall time, $t_{\rm ff}$.  

The cloud in Calculation 2 is more Jeans unstable than that of Calculation 1 (i.e.\ gravity dominates more over thermal pressure) and also has a higher initial turbulent Mach number.  These factors lead to several differences between Calculations 1 and 2.  Whereas star formation began at $t=1.037t_{\rm ff}=1.97\times 10^5~{\rm yr}$ in Calculation 1, star formation begins slightly earlier in Calculation 2 at $t=0.824t_{\rm ff}=5.22\times 10^4~{\rm yr}$.  The time-scales in years differ by a large factor because the second cloud is nine times denser and, hence, its free-fall time is shorter by a fact of three in real terms.  The rms Mach number of the turbulence has fallen from its initial value of ${\cal M}=9.2$ to ${\cal M}=5.6$ when the star formation begins in Calculation 2.  Note that this is still very high; such a dense cloud with a high-velocity dispersion is not typical of nearby star-forming regions, but in order to investigate the dependence of star formation on the mean thermal Jeans mass we are forced numerically to study very dense clouds as discussed in Section 2.3.  A second difference is that finer structures are visible in the gas in Calculation 2 (compare the last panel of Figure \ref{global} with the equivalent panel of Figure 2 in BBB2003).  This is due to the ability of thermal pressure to smooth out more massive structures in Calculation 1.

In Calculation 2, the regions of overdensity eventually evolve into four dense star-forming cores.  The properties of these cores and the numbers of stars and brown dwarfs they produce during the calculation are summarized in Table 2.  Calculation 1 produced three star-forming cores.  Because the structure of the initial turbulence was identical in the two calculations, the locations and masses of the cores in the calculations are similar.  The main dense core, Core 1, has a similar location and mass in both calculations.  Core 2 in Calculation 2 has roughly the same location and mass as Cores 2 and 3 in Calculation 1.  Two regions that did not form stars in Calculation 1 form stars in Calculation 2 because gravity is more dominant over thermal pressure in the second calculation.

In both calculations, the most massive dense core begins forming stars first (Figure \ref{core1}).  The four dense cores in Calculation 2 begin forming stars at $t=0.824t_{\rm ff}$, $t=1.111t_{\rm ff}$, $t=1.163t_{\rm ff}$, and $t=1.388t_{\rm ff}$, respectively.  Their gas masses just before they form their first objects and also when the calculation is stopped (at $t=1.40 t_{\rm ff}$) are given in Table \ref{table2} for comparison with Table 1 in BBB2003. The mass of a core is calculated as the amount of gas with density greater than some threshold value.  The masses of the cores depend on the density threshold that is used; because the mean density of the cloud has been increased by a factor of nine over Calculation 1, the two density thresholds used have been increase by an order of magnitude over those in BBB2003.

Both calculations were stopped at $t=1.40t_{\rm ff}$ to allow a direct comparison of the results.  Star formation would continue in both clouds if the calculations were followed further.  However, at this point there are a sufficient number of objects in both calculations that have reached their final masses to allow a meaningful comparison (Table \ref{table1}).  Calculation 2 produces 19 stars and 46 brown dwarfs.  An additional 14 objects have substellar masses but are still accreting.  Two of these have very low masses and accretion rates and therefore would probably end up with substellar masses if the calculation were continued.  Seven already have masses near the stellar/substellar boundary and are therefore likely to become stars.  The remaining five formed shortly before the simulation was stopped and it is impossible to tell whether they would become stars or not.

\begin{figure}
\centerline{\psfig{figure=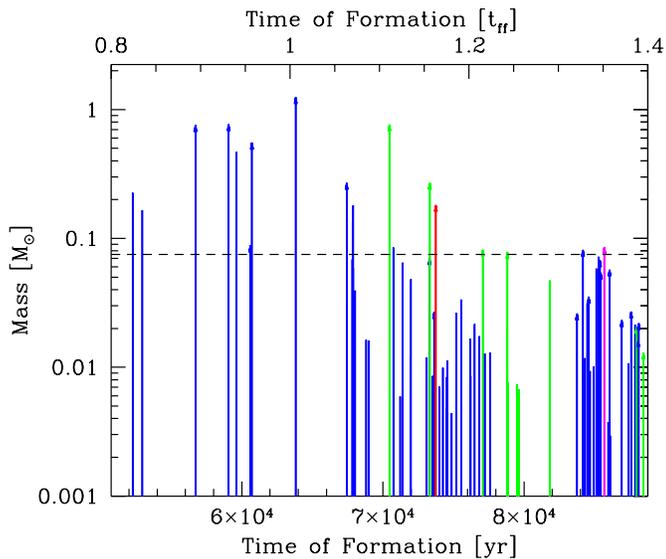,width=9.0truecm}}
\caption{\label{sfrate} Time of formation and mass of each star and brown dwarf 
at the end of Calculation 2.  The colour of each line identifies the dense core 
in which the object formed: first (blue), second (green), third (red), or fourth (magenta) core.
Objects that are still accreting significantly at the end of the calculation a
represented with arrows.  The horizontal dashed line marks the star/brown dwarf 
boundary.  Time is measured from the beginning of the calculation in terms of
the free-fall time of the initial cloud (top) or years (bottom).  This figure may be compared
with the equivalent figure for Calculation 1 contained in BBB2003.} 
\end{figure}

\subsection{The star-formation process in the dense cores}
\label{process}

Snapshots of the process of star formation in Cores 1 and 2 of Calculation 2 are shown in Figures \ref{core1} and \ref{core2}, respectively.  Cores 3 and 4 only produce 1 star each and are not shown.  As with Calculation 1, a true appreciation of how dynamic and chaotic the star-formation process is can only be obtained by studying an animation of the simulation.  The reader is encouraged to download an animation comparing Calculations 1 and 2 from http://www.astro.ex.ac.uk/people/mbate/Research/Cluster.

As in Calculation 1, the star formation in the two most massive cores of Calculation 2 proceeds via gravitational collapse to produce filamentary structures that fragment (e.g. Bastien 1983; Bastien et al.\ 1991; Inutsuka \& Miyama 1992) to form a combination of single objects and multiple systems (Figures \ref{core1} and \ref{core2}).  Many of the multiple systems result from the fragmentation of massive circumstellar discs (e.g. Bonnell 1994; Bonnell \& Bate 1994; Whitworth et al.\ 1995; Burkert, Bate \& Bodenheimer 1997; Hennebelle et al.\ 2004).  In the most massive core, most of the objects fall together into the gravitational potential well of the core to form a small stellar cluster (Figure \ref{core1}, $t=1.16-1.20 t_{\rm ff}$).  At this point, the cluster contains $\approx 39$ objects and is very compact so that dynamical interactions result in around half of the objects being ejected from the cluster in a short space of time.   More objects form as these objects are being ejected, but there is a brief pause in the star formation in core 1 from $t=1.23-1.32t_{\rm ff}$ (see also Figure \ref{sfrate}) because so much of the gas is used up in the star formation that more gas needs to fall into the potential well before another burst of star formation can occur.  The second burst lasts until the calculation is stopped.  Objects are again ejected from the cluster during this burst, but when the calculation is stopped the main dense core still contains $\approx 23$ objects.  In Calculation 1, it was also found that the star formation in the most massive dense core proceeded in bursts.  The main difference between the star formation occuring in the most massive cores of Calculations 1 and 2 is that in Calculation 2, the stellar group that formed is denser and more numerous leading to more violent dynamical interactions and fewer large discs (compare Figure \ref{core1} with Figures 3 and 4 of BBB2003).  
The dense cores formed in the calculations are formed by converging flows of gas.  Since the same turbulent flow structure was used for the initial conditions of both calculations, scaled to a smaller cloud for Calculation 2, it follows that the main dense core is more compact in Calculation 2 and, thus, also more Jeans unstable.  The implications of this more compact star-forming core will be discussed further in later sections.

Core 2 produces 12 objects during Calculation 2.  Six are formed directly through the fragmentation of filaments while the other six are formed through disc fragmentation events.  As in Calculation 1, dynamical interactions within these smaller groups work to arrange dynamically unstable systems into more stable configurations.  Three of the 12 objects are dynamically ejected during the calculation, but more ejections would be expected in the long term.

\subsection{Star formation timescale and efficiency}
\label{efficiency}

The timescale on which star formation occurs is the dynamical one in both calculations, consistent both with observational and other theoretical arguments (Pringle 1989; Elmegreen 2000a; Hartmann, Ballesteros-Paredes \& Bergin 2001), whether or not magnetic fields are present (Ostriker et al.\ 2001; Li et al.\ 2004).  However, comparing the two calculations, it is interesting to note that reducing the mean thermal Jeans mass (i.e. the degree of pressure support in the cloud) and increasing the turbulent Mach number leads to more rapid star formation.  Naturally, the star formation timescale in years is much shorter in Calculation 2 because it is a denser cloud and the free-fall time-scale is shorter.  However, even in terms of the number of initial cloud free-fall times, the star formation is more rapid. Calculation 1 converts 5.89 M$_\odot$ (12 percent) of its gas into stars in $1.40t_{\rm ff}$, while in the same number of free-fall times, Calculation 2 converts 7.92 M$_\odot$ (16 percent) of its gas into stars -- an increase of 34 percent.

In both calculations, the local star-formation efficiency is high within each of the dense cores (Table 2; for Calculation 1 see Table 1 of BBB2003).  This high star-formation efficiency is responsible for the bursts of star formation seen in both calculations.  Gas is rapidly converted into stars in the most massive dense cores and depleted to such an extent that star formation pauses.  Fresh gas must fall into the gravitational potential wells of the small clusters before new bursts of star formation can ensue.  Although the local star-formation efficiency is high in the dense cores, most of the gas in both calculations is in low-density regions where no star formation occurs.  Thus, the overall star formation efficiencies are low ($\sim 10$\%) for both calculations.  Although neither calculation has been followed until star formation ceases, in both calculations a large fraction of the gas has drifted off to large distances by the end of the calculation due to the initial velocity dispersion and pressure gradients and is not gravitationally unstable.  Thus, the global star formation efficiency is unlikely to exceed a few tens of percent.  We note that it is reasonable to assume that if the calculations were run until star formation ceased, Calculation 2 would have a higher overall star formation efficiency because thermal pressure gradients would be less important in driving low-density gas away from the cloud.  
Another aspect that impacts on the overall star formation efficiency is the initial magnitude of the turbulence.  Our initial conditions set the total turbulent kinetic energy to equal the magnitude of the gravitational energy.  Accounting for the small amounts of gas thermal energy, the clouds start off slightly unbound. Clark \& Bonnell \shortcite{ClaBon2004} recently studied star formation in clouds that were turbulently unbound initially.  They found that such initial conditions can result in very low star formation efficiencies.  Conversely, initial conditions with less turbulent support initially would be expected to give greater overall star formation efficiencies. 
Finally, although these calculations only form low-mass stars, feedback from jets, outflows and heating of the gas (none of which are included) would be expected to reduce the star formation efficiency further.

As discussed in BBB2003, observations show that star formation efficiencies vary widely across star-forming regions.  Some parts of star-forming clouds contain no newly formed objects while in other parts, notably clusters and groups, the local efficiency can reach 50\% or more.  Overall, such a pattern results in low global star formation efficiencies, typically 10-30\% (Wilking \& Lada 1983; Lada 1992).

\begin{figure}
\centerline{\psfig{figure=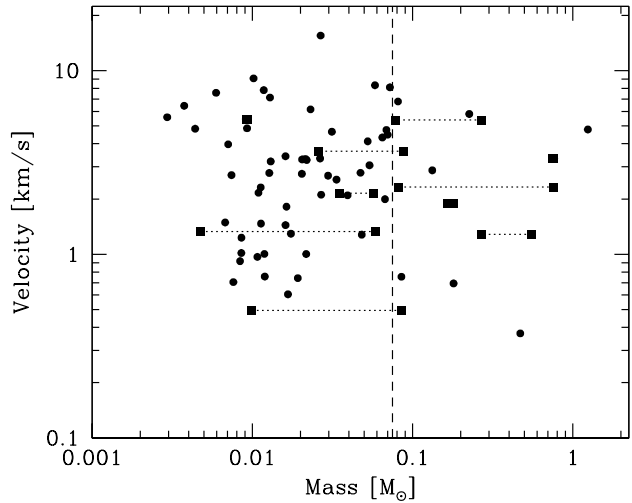,width=9.0truecm}}
\caption{\label{veldisp}  The velocities of each star and brown dwarf relative to
the centre-of-mass velocity of the stellar system.  For close binaries (semi-major 
axes $< 10$ AU), the centre-of-mass velocity of the binary is given, and the
two stars are connected by dotted lines and plotted as squares rather than circles.  
The root mean square velocity dispersion 
for the association (counting each binary once) is 4.3 km/s (3-D) or 2.5 km/s (1-D),
which is roughly a factor of two higher than in Calculation 1 due to the increased
densities involved.  As in Calculation 1, there is no significant dependence of the 
velocity dispersion on either mass or binarity.  
The vertical dashed line marks the star/brown dwarf boundary.} 
\end{figure}

\subsection{Stellar velocity dispersion and distribution}

As mentioned above, the small clusters formed in both calculations rapidly dissolve due to dynamical interactions between cluster members ejecting both stars and brown dwarfs.  Somewhat surprisingly, BBB2003 found that the final velocity dispersion of the stars and brown dwarfs is independent both of stellar mass and binarity.  While the lack of dependence on mass was reported from past $N$-body simulations of the breakup of small-$N$ clusters with $N>3$ (Sterzik \& Durisen 1998) and SPH calculations of $N=5$ clusters embedded in gas (Delgado-Donate, Clarke \& Bate 2003), these calculations found that binaries should have a smaller velocity dispersion than single objects due to the recoil velocities of binaries being lower, keeping them within the stellar groups.  

Calculation 2 gives similar results to Calculation 1 in that there is no statistically significant difference in the velocity dispersion of brown dwarfs versus stars, or of singles versus binaries (Figure \ref{veldisp}).  
The lack of dependence of the velocity dispersion on binarity is due to three reasons.  First, the presence of gaseous discs surrounding objects formed in the calculations discussed here allows disc fragmentation and dissipative interactions between stars, both of which create tight binaries (see Bate et al.\ 2002b).  These tight binaries can easily be ejected in dynamical interactions.  In the $N$-body/SPH simulations of Delgado-Donate et al.\ (2003), discs were absent and the probability of forming and ejecting a tight binary, especially given the small numbers of objects in the groups, was negligible.  Typically, only single objects were ejected leaving a binary behind.  More recent small-$N$ SPH cluster simulations by Delgado-Donate et al. (2004a) that begin from turbulent initial conditions which allow circumstellar disc formation and fragmentation also show that the velocity dispersions of singles and binaries are similar, supporting this hypothesis.  The second reason for the difference between the above simulations and the calculations reported here is that the stellar velocities are contributed to by the motions of individual dense cores (BBB2003; Goodwin, Whitworth \& Ward-Thompson 2004).  This source of velocity dispersion was not considered in the above studies because each calculation modelled only a single core.
The third reason is due to something that did not occur in Calculation 1.  In Calculation 2, some wide binaries are formed when two ejected objects happen to be ejected at similar times and with similar velocities and, thus, are gravitationally bound.  These are ejected binaries, so have speeds of a few km/s, but are {\it not ejected as binaries}.  These systems are discussed more in Section \ref{multiplesystems}.

The velocities of the stars and brown dwarfs relative to the centre of mass of all the objects are given in Figure \ref{veldisp}.  The rms velocity dispersion is 4.3 km~s$^{-1}$ in three dimensions or 2.5 km~s$^{-1}$ in one dimension (using the centre-of-mass velocity for binaries closer than 10 AU).  This is roughly a factor of 4 greater than the three-dimensional (3D) velocity dispersion of the gas when the stars begin to form (${\cal M}=5.6$ giving a 3D velocity dispersion of 1.0 km~s$^{-1}$).  Thus, dynamical interactions are the primary source of the overall velocity dispersion.  Comparing the magnitude of the velocity dispersions from the two calculations, the value from Calculation 2 is a factor of two higher than that for Calculation 1 (which was 2.1 km~s$^{-1}$).  A higher velocity dispersion is to be expected in Calculation 2 simply on the grounds that the cloud is a factor of 2.08 smaller in radius and thus virial arguments would imply an increase in the velocity dispersion by a factor of $\approx \sqrt{2}$.  
However, the increase is somewhat greater than this factor.  This is probably due to the fact that, as noted in Section \ref{process}, the cluster formed in the main dense core of Calculation 2 collapses to a very compact state just before many of the objects are ejected.

Observationally, in agreement with the calculations presented here, there is no evidence for brown dwarfs having a significantly higher velocity dispersion than stars (something that was suggested as a possible signature that brown dwarfs form as ejected stellar embryos by Reipurth \& Clarke 2001).  In fact, Joergens \& Guenther (2001) studied the radial velocities of stars and brown dwarfs in the Chamaeleon I dark cloud and found the brown dwarfs had a velocity dispersion of $\approx 2$ km~s$^{-1}$ while the overall velocity dispersion was $\approx 3.6$ km~s$^{-1}$.  It is thought that overall value is high due to the radial velocity `noise' exhibited by T Tauri stars (Guenther et al.\ 2001).  An increase in velocity dispersion with the density of a star-forming region is suggested observationally.  The one-dimensional velocity dispersion in Taurus-Auriga has been measured at $\lsim 2$ km~s$^{-1}$ using proper motions (Jones \& Herbig 1979; Hartmann et al.\ 1991; Frink et al.\ 1997), while in the dense Orion Trapezium cluster, the 1D velocity dispersion is $2.3$ km s$^{-1}$ (Jones \& Walker 1988; Tian et al.\ 1996).

\begin{figure*}
\centerline{\psfig{figure=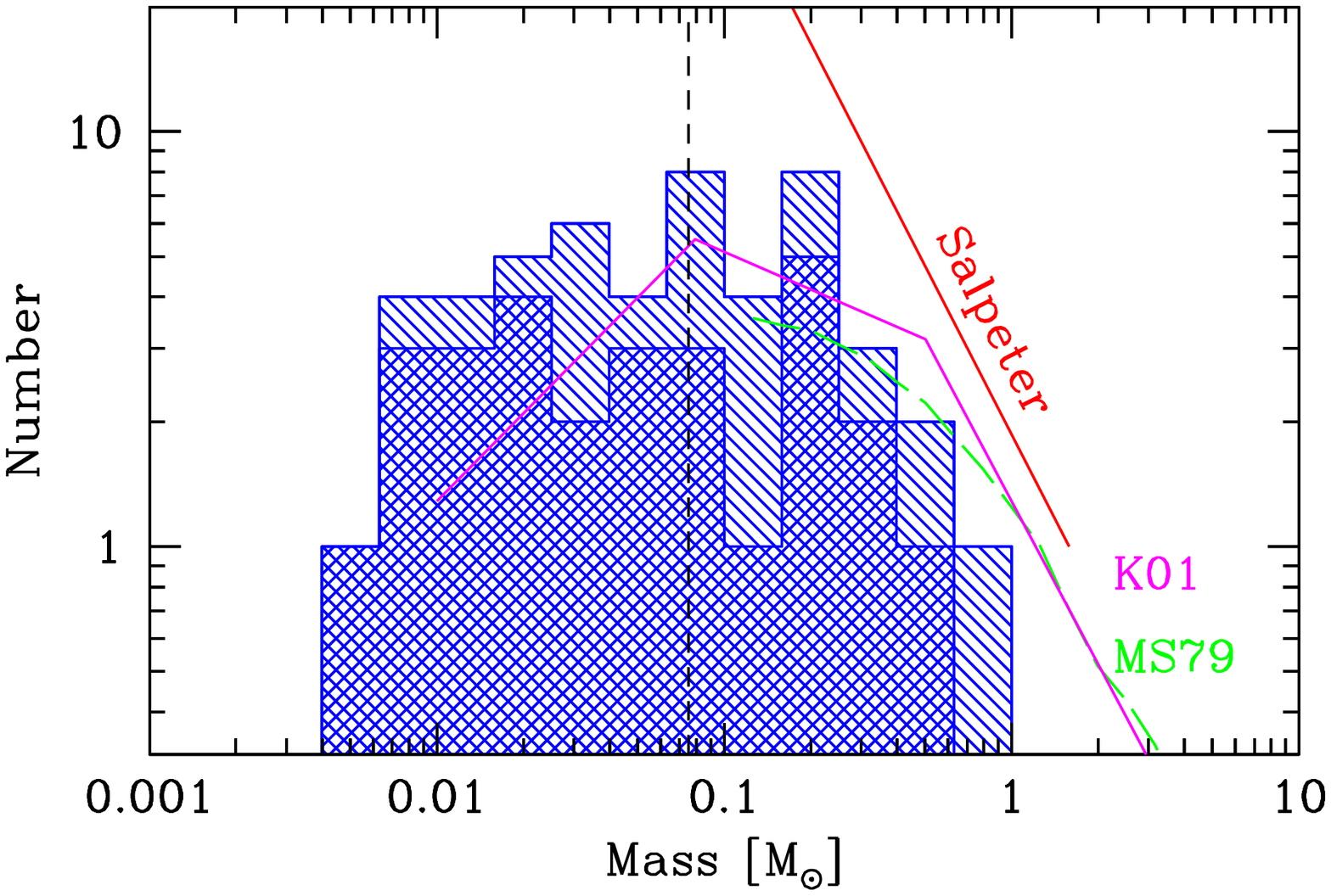,width=9.0truecm}\psfig{figure=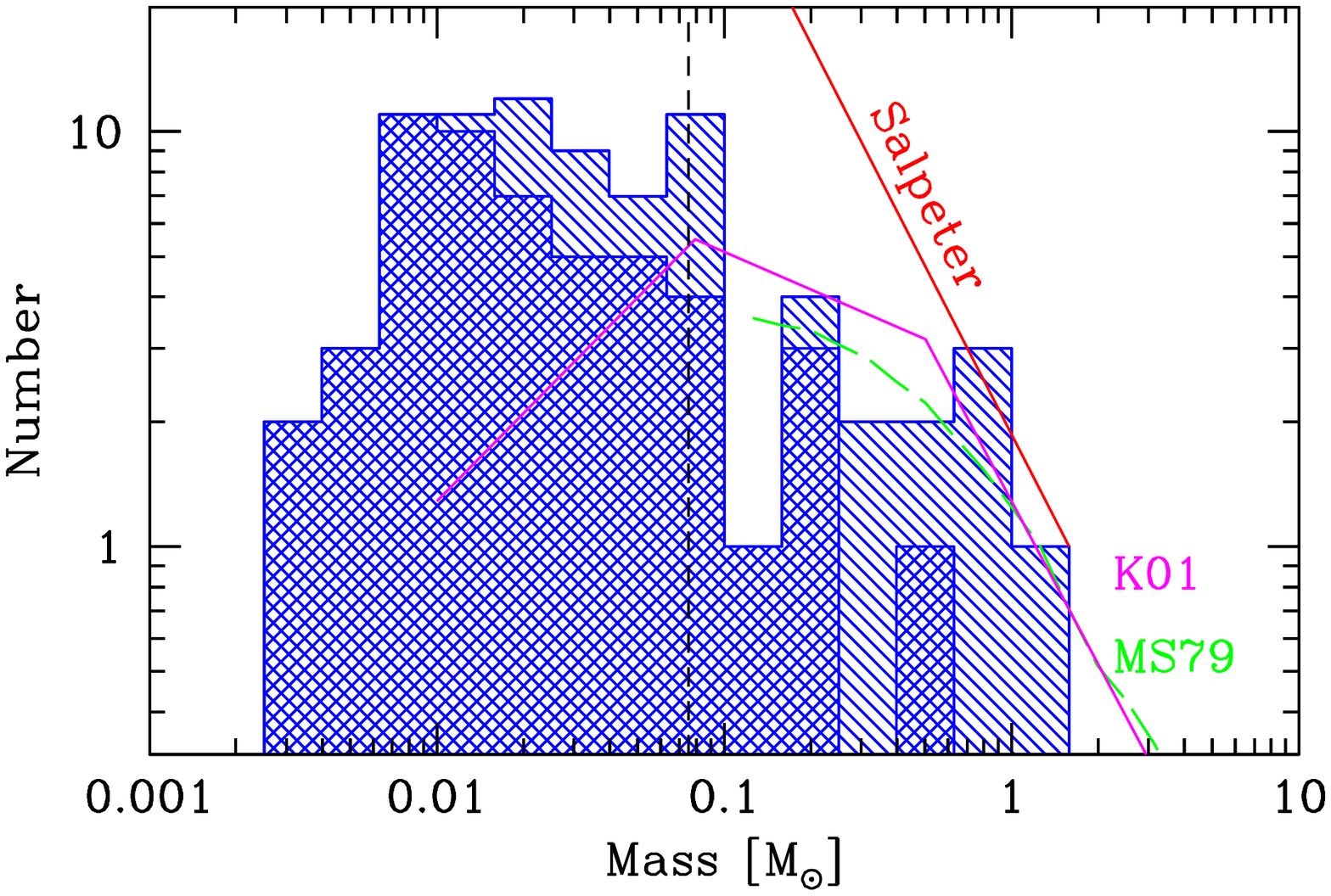,width=9.0truecm}}
\caption{\label{imf} The initial mass functions produced by the two calculations.  
Calculation 2 (righthand panel) had the lower initial mean thermal Jeans mass and 
produced a much higher fraction of brown dwarfs.
The single shaded regions show all of the objects, the double shaded regions 
show only those objects that have finished accreting.  The mass resolution of the 
simulations is 0.0011 M$_\odot$ (i.e.\ 1.1 M$_{\rm J}$), but no objects have 
masses lower than $2.9$ M$_{\rm J}$ due to the opacity limit for fragmentation.
We also plot fits to the observed IMF from Miller \& Scalo (1979) (dashed line) 
and Kroupa (2001) (solid broken line).  The Salpeter (1955) slope (solid straight line) is equal to that of 
Kroupa (2001) for $M>0.5$ M$_\odot$.  The vertical dashed line marks
the star/brown dwarf boundary.} 
\end{figure*}

\subsection{Initial mass function}
\label{imfsec}

A summary of the mass distributions of the stars and brown dwarfs formed in two calculations is given in Table 1.  Calculation 1 formed roughly equal numbers of stars and brown dwarfs.  The mean object mass was 0.118 M$_\odot$ and the median mass was 0.070 M$_\odot$.  Calculation 2 formed more objects (79 versus 50) in the same number of initial cloud free-fall times.  Two thirds to three quarters of these objects are brown dwarfs (depending on whether you include all substellar objects, or only those that have stopped accreting).  The mean object mass was 0.100 M$_\odot$ and the median mass was only 0.023 M$_\odot$.

In Figure \ref{imf} we plot the initial mass functions obtained from the two calculations, covering both the stellar and sub-stellar regimes.  The mimimum resolvable mass in the calculations is 1.1 M$_{\rm J}$, but there are no objects formed with masses this low due to the opacity limit for fragmentation.  The lowest object mass was 5 M$_{\rm J}$ in the first calculation and 3 M$_{\rm J}$ in the second.  The precise value of the cutoff in the IMF is not well constrained by these calculations because the opacity limit for fragmentation is modelled using the equation of state in equation \ref{eta} rather than by performing radiative transfer.  However, the above masses are consistent with the predicted cutoff at $\sim 1-10$ M$_{\rm J}$ (Low \& Lynden-Bell 1976; Rees 1976; Silk 1977; Boss 1988; Masunaga \& Inutsuka 1999; Boss 2001).

The initial mass function (IMF) obtained from the first calculation is consistent with
\begin{equation}
\frac{{\rm d}N}{{\rm d log}M} \propto M^{\Gamma}
\end{equation}
where
\begin{equation}
\Gamma = \left\{ \begin{array}{rl}
                -1.35 &\mbox{\rm for $M\gsim 0.5$~M$_\odot$} \\
                0.0 & \mbox{\rm for $0.005 < M \lsim 0.5$~M$_\odot$}
               \end{array}
         \right.
\end{equation}
and there are no objects below the opacity limit for fragmentation
($\approx 0.005$~M$_\odot$).
The Salpeter slope is $\Gamma=-1.35$ \cite{Salpeter1955}.  In turn, this is broadly consistent with the observed IMF (e.g. Luhman et al.\ 2000; Kroupa 2001; Chabrier 2003).

The {\it stellar} IMF produced by the second calculation is again broadly consistent with the observed IMF, however, as already seen in Table 1, the second calculation produces many more brown dwarfs.  Overall, this IMF is consistent with 
\begin{equation}
\Gamma = \left\{ \begin{array}{rl}
                -1.35 &\mbox{\rm for $M\gsim 0.5$~M$_\odot$} \\
                -0.3 & \mbox{\rm for $0.005 < M \lsim 0.5$~M$_\odot$}.
               \end{array}
         \right.
\end{equation}
The cut-off at the opacity limit for fragmentation is again very steep, but there are three objects with final masses slightly lower than 0.005 M$_\odot$ (2.9, 3.8, and 4.4 Jupiter-masses).  Note that in neither calculation is the slope above 0.5 \msun\ well constrained.

\begin{figure}
\centerline{\psfig{figure=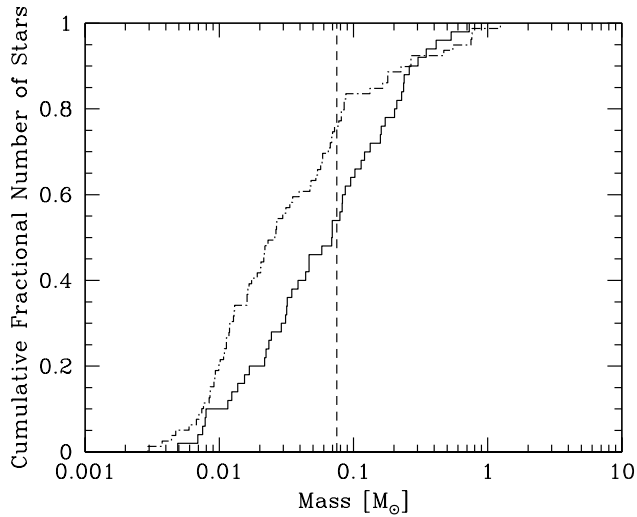,width=9.0truecm}}
\caption{\label{cumimf} The {\it cummulative} initial mass functions produced 
by Calculations 1 (solid line) and 2 (dotted line).  Again, the excess of brown dwarfs
in the second calculation over the first is clear.  A Kolmogorov-Smirnov test on the
two distributions shows that there is only a 1.9\% probability that they are drawn
from the same underlying IMF.  
The vertical dashed line marks the star/brown dwarf boundary.} 
\end{figure}

Despite the use of large-scale hydrodynamical calculations, the accuracy with which we can determine the resulting IMFs is limited by small number statistics.  Thus, we must ask whether or not the two distributions are in fact significantly different.  In Figure \ref{cumimf}, we give the cumulative IMFs from the two calculations.  A Kolmogorov-Smirnov test performed on the distributions tells us that there is only a 1.9 percent probability that they are drawn from the same underlying IMF (i.e. they differ at the 98.1 percent confidence level; roughly a $2.4\sigma$ result).  The difference between the two IMFs is again clear from Figure \ref{cumimf}; there are more brown dwarfs formed in the calculation with the lower initial thermal Jeans mass.

This result is, perhaps, not too surprising.  Larson \shortcite{Larson1992}, for example, proposed that the characteristic stellar mass should be linked to the Jeans mass in molecular clouds.  However, Table 1 shows that it is not quite that simple.  While reducing the thermal Jeans mass by a factor of 3 gives a corresponding drop in the median object mass by a factor of 3.04 from 0.070 $M_\odot$ to 0.023 M$_\odot$, the mean object mass decreases by only 17 percent.  Why does the median object mass depend on the thermal Jeans mass, while the mean mass does not? This will be addressed in Section \ref{variation}.  Note that it has also been argued that the characteristic stellar mass may be related to the turbulent Jeans mass, rather than the thermal Jeans mass.  The initial Mach number is 40\% greater in Calculation 2 than Calculation 1 (50\% greater when the first star forms in each calculation).  Thus, the initial turbulent Jeans mass differs by a factor of $3\times 1.4=4.2$ between the two calculations.  Neither the mean nor the median stellar mass change by a factor this great.

\subsection{The abundance of brown dwarfs}

In this section, we investigate in detail why the second calculation forms more brown dwarfs than the first.  From Calculation 1, Bate et al. (2002a) found that the mechanism for brown dwarf formation was that a fragment was ejected from the region of dense molecular gas in which it formed before it was able to accrete to a stellar mass.  The ejections occurred due to dynamical interactions in unstable multiple systems.  This brown dwarf formation mechanism was proposed by Reipurth \& Clarke (2001).  Thus, the formation of more brown dwarfs in the second calculation implies either that the accretion rates on to the fragments were lower, or that the objects were ejected more quickly after they formed.

\begin{figure*}
\centerline{\psfig{figure=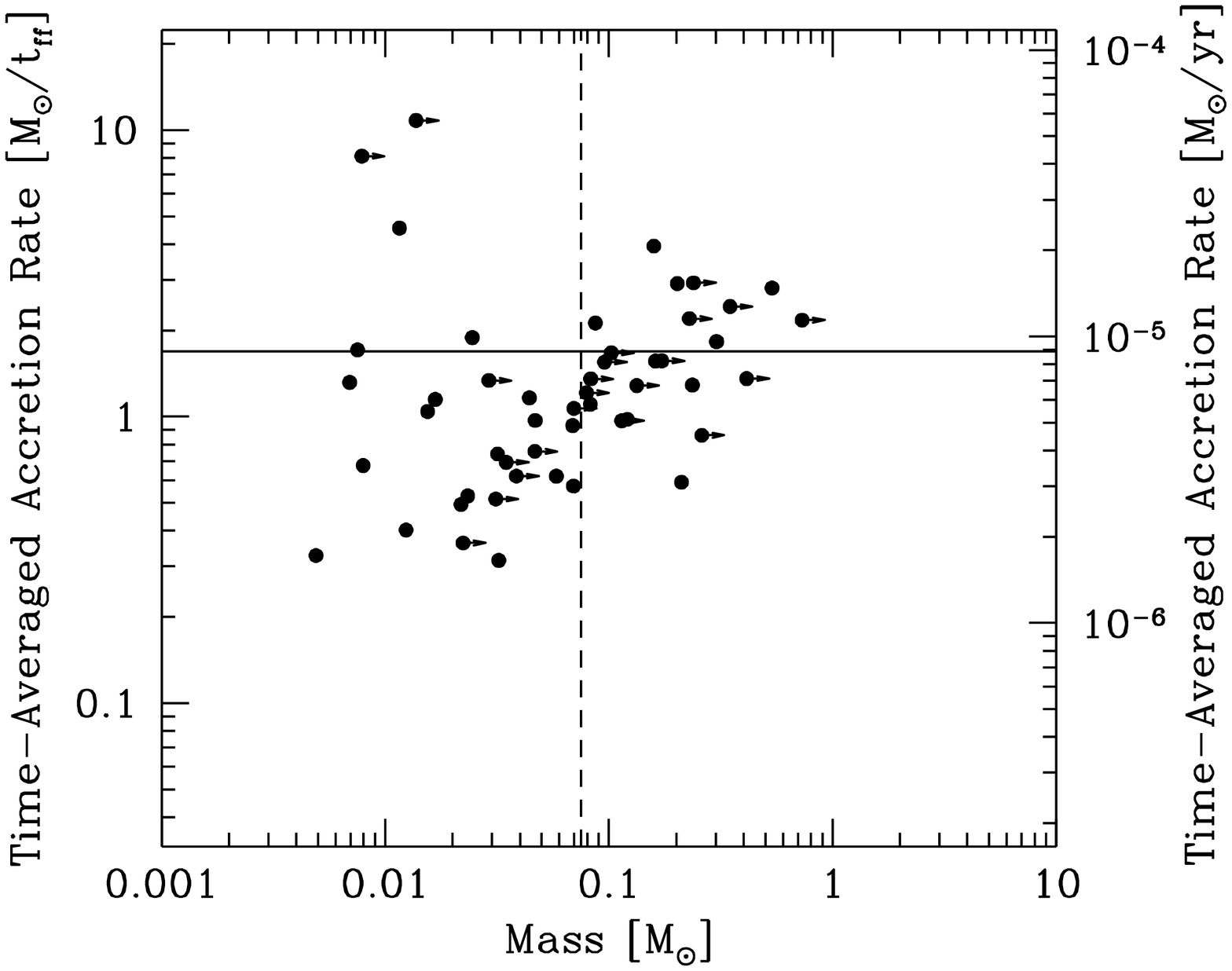,width=7.0truecm}\hspace{1.0cm}\psfig{figure=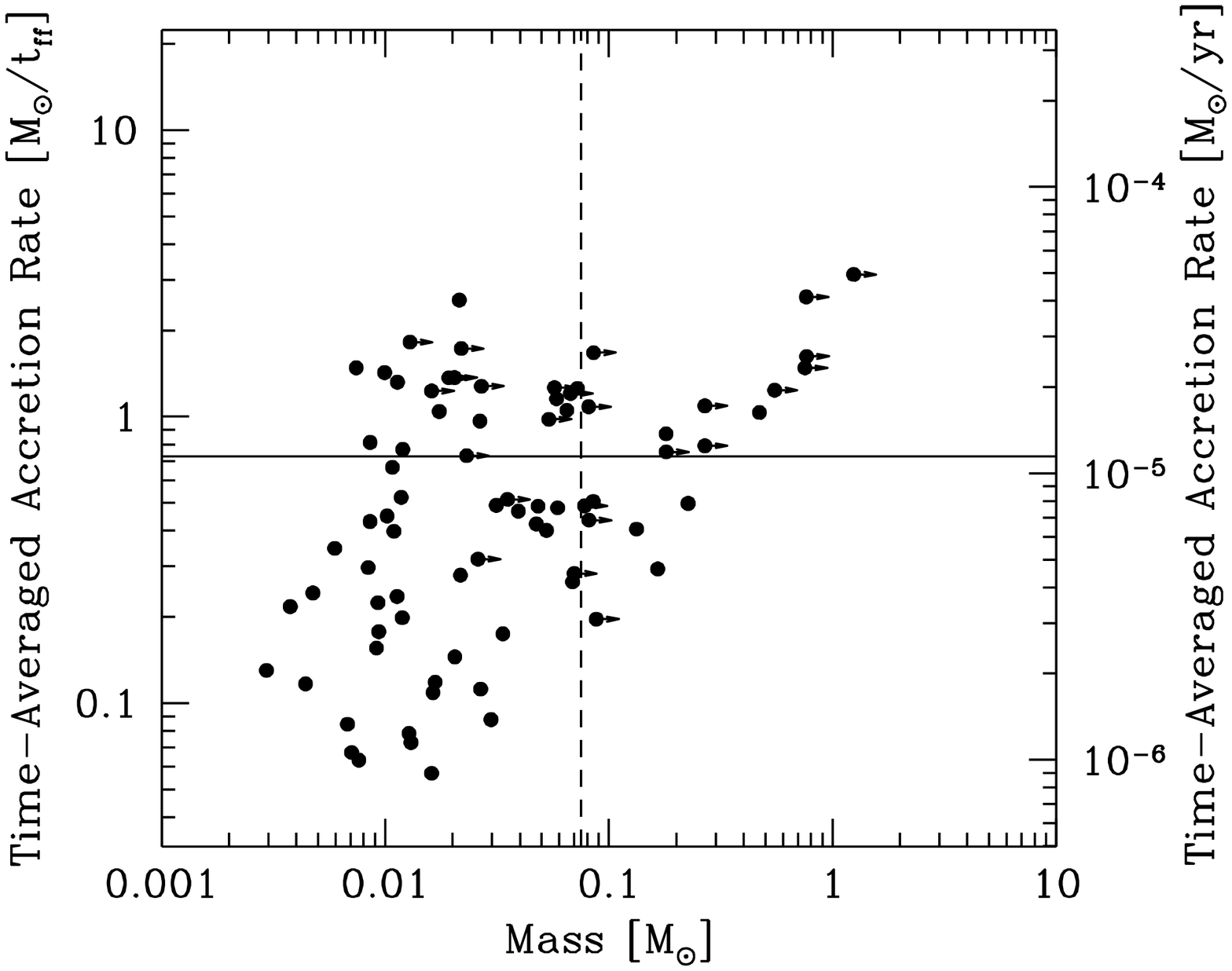,width=7.0truecm}}
\caption{\label{accrates} The time-averaged accretion rates of the objects formed in the two
calculations versus their final masses.  The accretion rates are calculated as the final mass of an object divided by the time between their formation and the termination of their accretion
or the end of the calculations.  The horizontal solid lines give the
means of the accretion rates: $8.6\times 10^{-6}$ M$_\odot$/yr  and $11.1\times 10^{-6}$ M$_\odot$/yr 
for Calculations 1
and 2, respectively.  The accretion rates are given in M$_\odot/t_{\rm ff}$ on the left-hand axes
and M$_\odot$/yr on the right-hand axes. The vertical dashed line marks the star/brown dwarf boundary.} 
\end{figure*}

In Figure \ref{accrates}, we plot the time-averaged accretion rates of all the objects for both Calculations 1 and 2.  A time-averaged accretion rate is defined as the mass of an object at the end of the calculation divided by the time over which it accreted that mass.  The accretion time is measured from the formation of an object (i.e., the insertion of a sink particle) to the {\it last} time at which its accretion rate drops below $10^{-7}$ M$_\odot$/yr, or the end of the calculation (which ever occurs first).  We also define an ejection time, which is the time between the formation of an object and {\it last} time the magnitude of its acceleration drops below 1000 km/s/Myr for Calculation 1 and 5000 km/s/Myr for Calculation 2 (or the end of the calculation).  The acceleration criterion is based on the fact that once an object is ejected from a stellar cluster through a dynamical encounter, its acceleration will drop to a low value.  The specific values of the acceleration mentioned above were chosen by comparing animations and graphs of acceleration versus time for individual objects.

It can be seen that the time-averaged accretion rates of the objects have a significant dispersion.  However, with the possible exception of the highest mass objects in Calculation 2, there is no systematic trend for the lower-mass objects to have lower time-averaged accretion rates.  Similarly, the accretion rates do not appear to be systematically lower in the second calculation.  The means of the time-averaged accretion rates are $8.6\times 10^{-6}$ and $11.1 \times 10^{-6}$ M$_\odot$/yr for Calculations 1 and 2, respectively.  Therefore, we conclude that the increased proportion of brown dwarfs in Calculation 2 is not due to lower accretion rates.  As a rough estimate, the mean accretion rates might be expected to depend on the sound speed (the same for both calculations) as $\sim c_{\rm s}^3/G=1.5\times 10^{-6}$ M$_\odot$/yr (Shu 1977; Hunter 1977).  The means of the time-averaged accretion rates are factors of a few higher than this estimate, but this is consistent with the fact that collapsing non-singular isothermal spheres usually accrete at a rate somewhat larger than $c_{\rm s}^3/G$ (e.g.\ Foster \& Chevalier 1993).

\begin{figure*}
\centerline{\psfig{figure=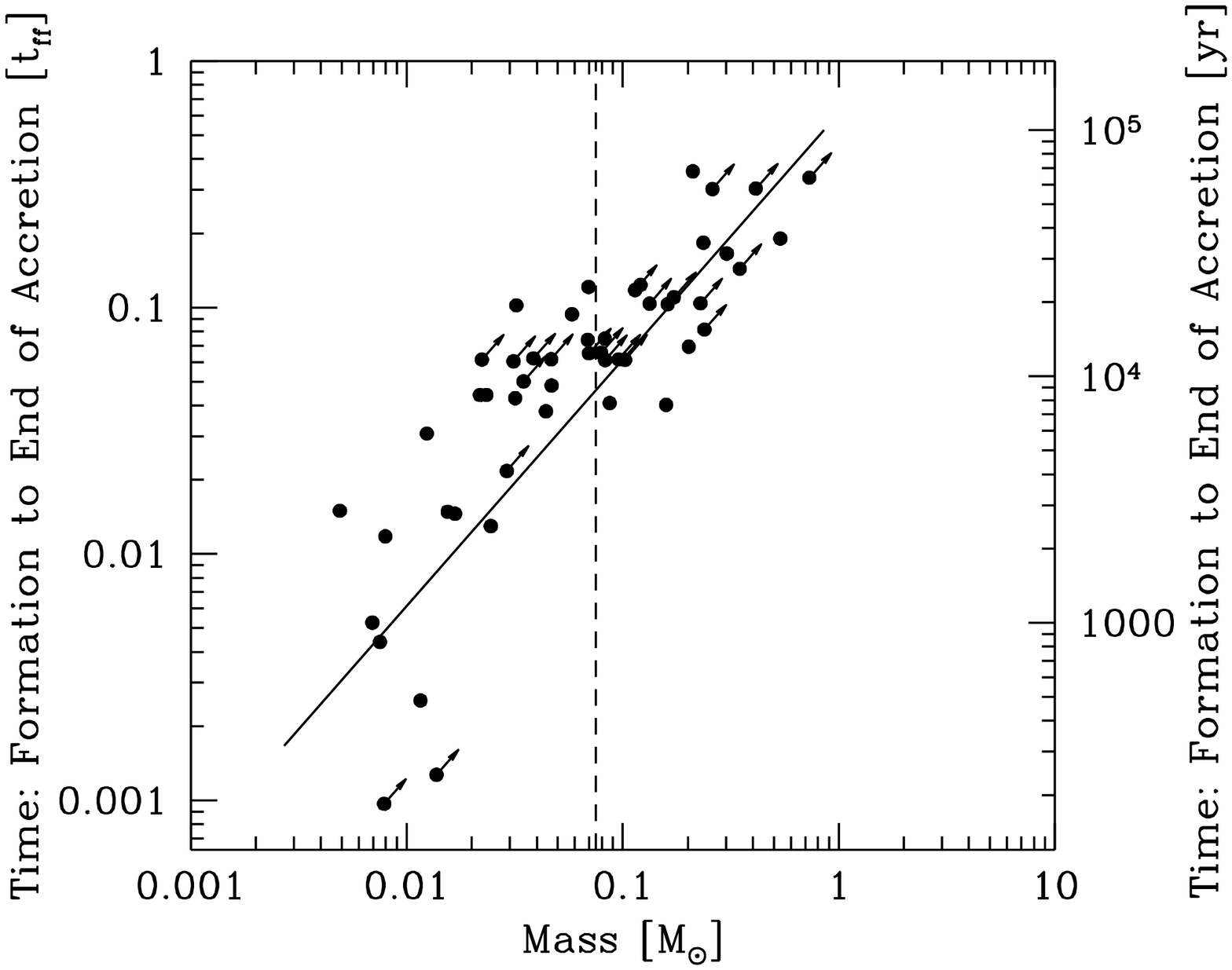,width=7.0truecm}\hspace{1.0cm}\psfig{figure=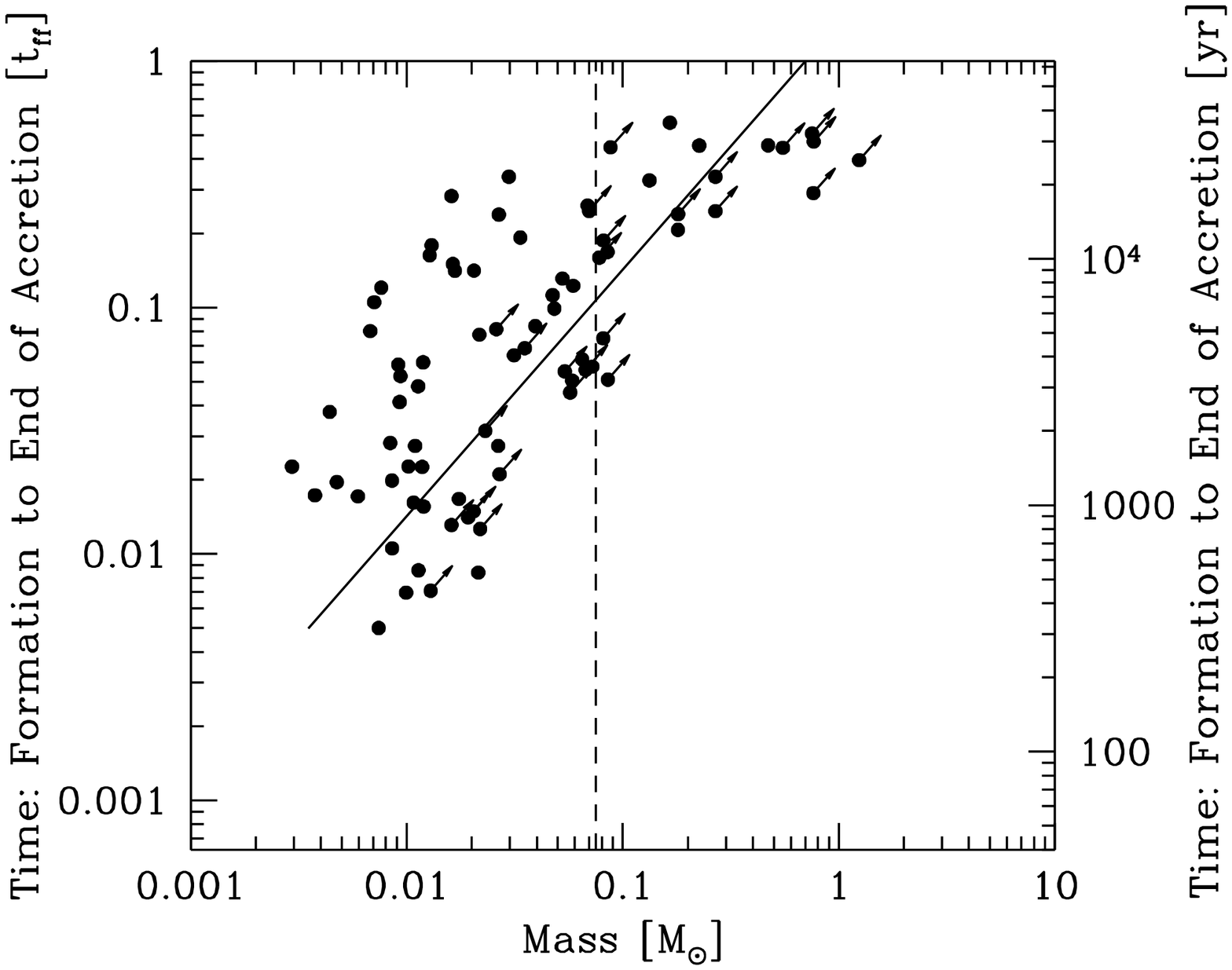,width=7.0truecm}}
\caption{\label{accmass} The time between the formation of each object and the termination
of its accretion or the end of the calculation versus its final mass.  In both calculations, there is 
a clear linear correlation between the time an object spends accreting and its final mass.  
The solid line gives the curve that the objects would lie on if each object accreted at the 
mean of the time-averaged accretion rates.
The accretion times are given in units of the $t_{\rm ff}$ on the left-hand axes
and years on the right-hand axes. The vertical dashed line marks the star/brown dwarf boundary.} 
\end{figure*}

In Figure \ref{accmass}, we plot the time between the formation of an object and the termination of its accretion (or the end of the calculation) versus the final mass of the object.  Those points with arrows denote those objects that are still accreting significantly at the end of the calculation.  Accreting objects would move towards the upper right of the diagrams if the calculations were extended.  From both calculations it is clear that the lower the final mass of the object, the earlier its accretion was terminated.  We also see that in the second calculation a much greater fraction of the objects have their accretion terminated soon after their formation (less than $10^4$ years).  This is the origin of the larger fraction of brown dwarfs in Calculation 2.  

\begin{figure*}
\centerline{\psfig{figure=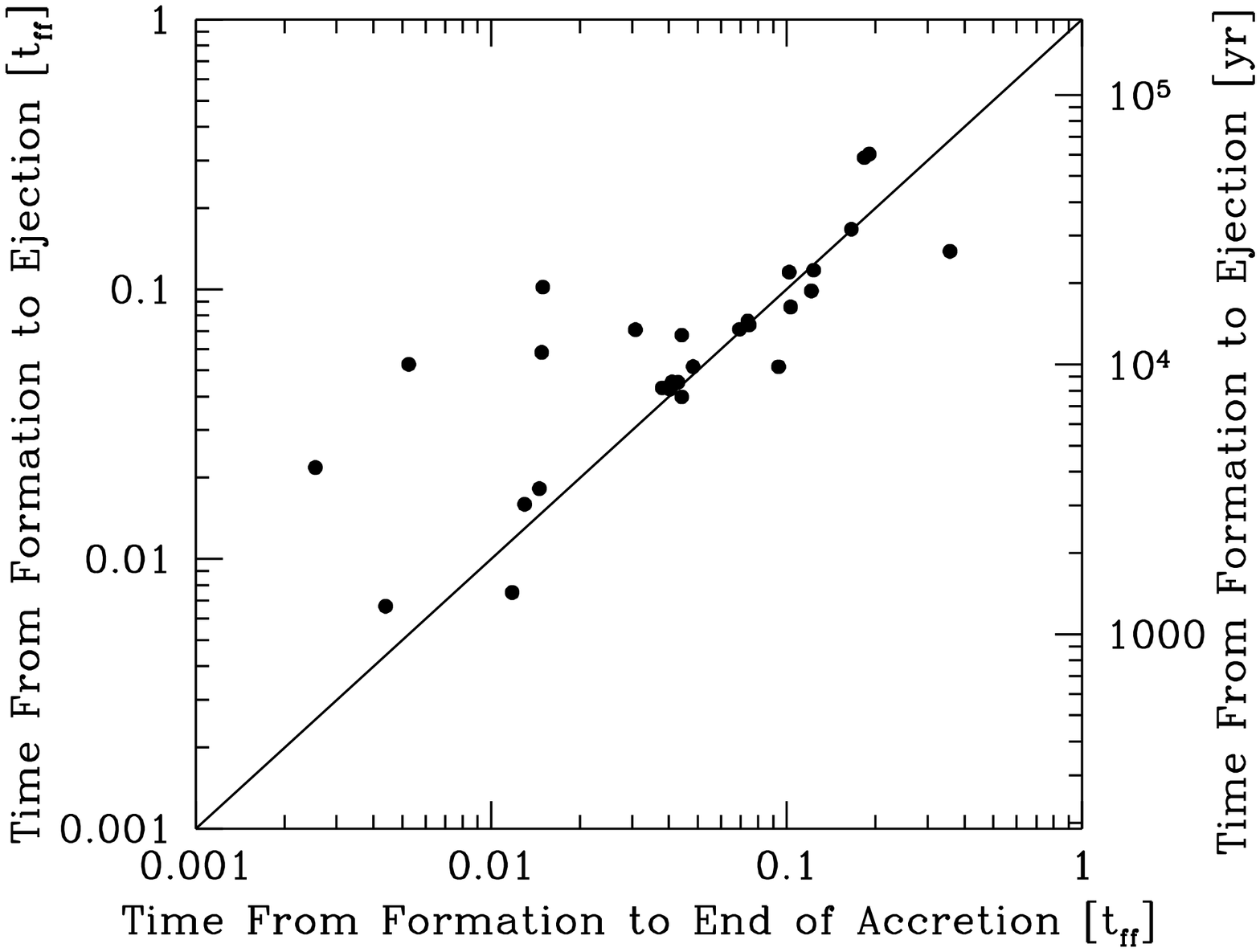,width=7.0truecm}\hspace{1.0cm}\psfig{figure=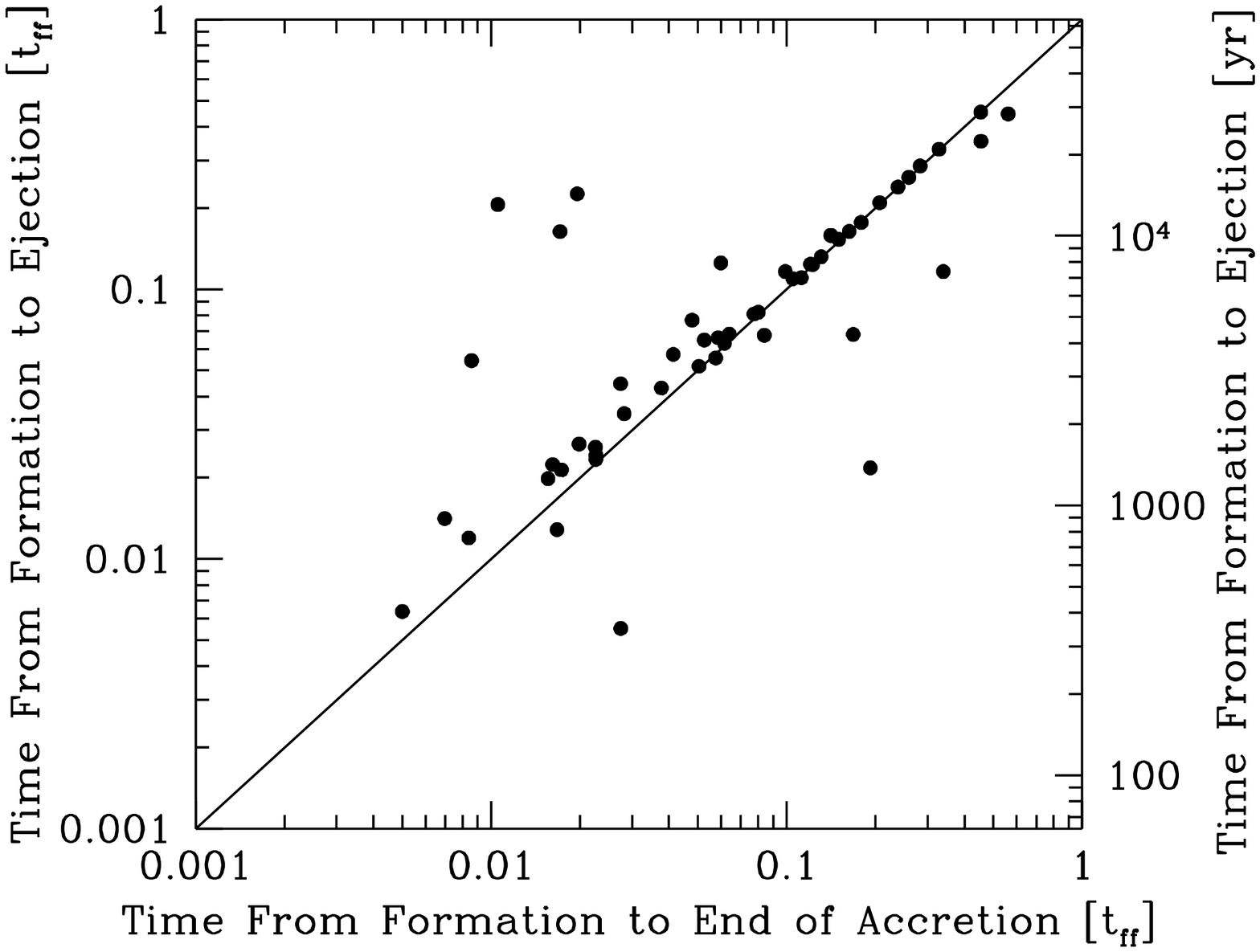,width=7.0truecm}}
\caption{\label{ejectacc} For each object that has stopped accreting, we plot the time between the formation of the object and its ejection from a multiple system versus the time between its formation 
and the termination of its accretion.  In both calculations, these times are correlated, showing that the termination of accretion on to an object is usually associated with dynamical ejection of the object. } 
\end{figure*}

What causes the termination of the accretion?  In Figure \ref{ejectacc}, we plot the time between the formation of an object and its ejection from a stellar group versus the time between the formation of an object and the termination of its accretion.  In this figure, we only plot those objects that have stopped accreting and reached their final masses by the end of the calculations.  In both calculations, these times are correlated, showing that {\it the termination of accretion on to a object is usually associated with dynamical ejection of the object}.  These results confirm the assertion of Bate et al. (2002a) that brown dwarfs are `failed stars'.  They fall short of reaching stellar masses because they are cut off from their source of accretion prematurely due to ejection in dynamical interactions.

Why are objects ejected more quickly in the second calculation than in the first?  The two calculations are identical except for the initial radii of the clouds.  This results in the second calculation being nine times denser and its initial mean thermal Jeans mass being three times lower.  Thus, the dynamical timescale of the second cloud is three times shorter and dynamical interactions tend to occur in a shorter real time in the second calculation.  On the other hand, as we have seen, the accretion rates of the objects are roughly the same for the two calculations (as is expected if the accretion rates scale with the sound speed of the gas).  Therefore, the reason for the larger proportion of low-mass objects in Calculation 2 is that the typical object accretes at the same rate as in Calculation 1, but does so for $\approx 1/3$ of the time (see also Section 4.2.3).

\begin{table*}
\begin{tabular}{lcccccl}\hline
Object Numbers & $M_1$ & $M_2$  & $q$ & $a$  & $e$  & Comments \\
        & M$_\odot$ & M$_\odot$ &  & &  \\ \hline
27,49    & 0.27  & 0.078  & 0.29  & 1.5*   & 0.86*  & Star+VLM\\ 
60,59    & 0.009  & 0.009  & 0.98  & 2.3*   & 0.93*  & BBD (accreting)\\ 
4,3   & 0.76  & 0.75  & 0.98  & 3.2*   & 0.77*  & \\ 
7,10   & 0.55  & 0.27 & 0.49  & 4.7*   & 0.30*  & \\
19.45   & 0.76  & 0.081 & 0.11  & 6.2*   & 0.42*  & Star+VLM, EMR\\
6,54   & 0.088 & 0.026 & 0.30  & 15   & 0.34  & VLM+BD (accreting)\\
69,58  & 0.057 & 0.035 & 0.62  & 21   & 0.59  & BBD (accreting)\\ 
15,33  & 0.059 & 0.005 & 0.081  & 66   & 0.82  & Ejected BBD, EMR \\
20,32  & 0.085 & 0.010 & 0.12  & 126   & 0.52  & Ejected VLM+BD, EMR \\ 
14,2  & 0.18 & 0.17 & 0.92  & 1136   & 0.98  & Ejected \\ \hline
(4,3),8 & (1.52) & 1.24 & 0.82 & 41    & 0.04  & \\
(7,10),26 & (0.82) & 0.070 & 0.085 & 45    & 0.67  & BD companion \\ \hline
(19,45),(27,49) &(0.84)&(0.35)&0.41&31  & 0.37  &  \\
((4,3),8),64 & (2.76) & 0.072 & 0.026 & 47 & 0.49 & BD companion \\
((7,10),26),41 & (0.89) & 0.017 & 0.019 & 407 & 0.90 & BD companion \\ \hline
(((4,3),8),64),71 & (2.83) & 0.023 & 0.008 & 58 & 0.34 & BD companion \\
((19,45),(27,49)),53 &(1.19)&0.047&0.040&118  & 0.33  & BD companion \\ \hline
((((4,3),8),64),71),72 & (2.85) & 0.011 & 0.004 & 72 & 0.97 & BD companion \\
(((19,45),(27,49)),53),79 &(1.24)&0.013&0.010&75  & 0.50  & BD companion \\ \hline
(((((4,3),8),64),71),72),57 & (2.87) & 0.031 & 0.011 & 168 & 0.48 & BD companion \\
((((19,45),(27,49)),53),79),75 &(1.25)&0.020&0.016&387  & 0.80  & BD companion \\ \hline
((((((4,3),8),64),71),72),57),78 & (2.90) & 0.022 & 0.008 & 203 & 0.23 & BD companion \\
(((((19,45),(27,49)),53),79),75),50 &(1.27)&0.008&0.006&489  & 0.95  & BD companion \\ \hline
(4,3),8),64),71),72),57),78),47 & (2.92) & 0.013 & 0.004 & 290 & 0.26 & BD companion \\
(4,3),8),64),71),72),57),78),47),63 & (2.93) & 0.009 & 0.003 & 309 & 0.85 & BD companion \\
(4,3),8),64),71),72),57),78),47),63),13 & (2.94) & 0.13 & 0.045 & 328 & 0.70 &  \\
(4,3),8),64),71),72),57),78),47),63),13),28 & (3.07) & 0.009 & 0.003 & 488 & 0.45 & BD companion  \\
(4,3),8),64),71),72),57),78),47),63),13),28),(60,59) & (3.08) & (0.018) & 0.006 & 491 & 0.89 & BBD companion  \\
(4,3),8),64),71),72),57),78),47),63),13),28),(60,59)),66 & (3.10) & 0.054 & 0.017 & 561 & 0.80 & BD companion  \\
(4,3),8),64),71),72),57),78),47),63),13),28),(60,59)),66),(6,54) & (3.15) & (0.11) & 0.036 & 570 & 0.69 & BBD companion  \\
(4,3),8),64),71),72),57),78),47),63),13),28),(60,59)),66),(6,54)),77 & (3.27) & 0.016 & 0.005 & 490 & 0.75 & BD companion  \\
(Above system),38 & (3.28) & 0.011 & 0.003 & 667 & 0.24 & BD companion  \\
(Above system),38),65 & (3.30) & 0.067 & 0.020 & 699 & 0.60 & BD companion  \\
(Above system),38),65),40 & (3.36) & 0.034 & 0.010 & 801 & 0.29 & BD companion  \\
(Above system),38),65),40),35 & (3.40) & 0.011 & 0.003 & 818 & 0.37 & BD companion  \\
(Above system),38),65),40),35),34 & (3.41) & 0.008 & 0.002 & 1012 & 0.68 & BD companion  \\
\hline
\end{tabular}
\caption{\label{table3} The properties of the 7 multiple systems with semi-major axes less than 1000 AU formed in the calculation (see also Figure 11).  Four of these systems are pure binaries while the other three have 4, 8, and 23 members.  The structure of each system is described using a binary hierarchy.  For each `binary' we give the masses of the primary $M_1$ and secondary $M_2$, the mass ratio $q=M_2/M_1$, the semi-major axis $a$, and the eccentricity $e$.  The combined masses of multiple systems are given in parentheses.  Orbital quantities marked with asterisks are unreliable because these close binaries have periastron distances less than the gravitational softening length.  When the calculation is stopped, the three high-order systems are unstable and/or are still accreting, so their final states are unknown.  Binary system (69,58) is also accreting. In the comments, BBD refers to a binary brown dwarf system, VLM refers to a very low-mass star (mass $<0.09$ M$_\odot$), EMR refers to an extreme mass ratio ($M_2/M_1<0.2$), and 'ejected' refers to binaries that have been ejected from the cloud.}
\end{table*}

\subsection{Multiple systems}
\label{multiplesystems}

As in Calculation 1, the dominant formation mechanism for binary and multiple systems in Calculation 2 is fragmentation, either of gaseous filaments (e.g.\ Bastein 1983; 
Bastien et al.\ 1991; Inutsuka \& Miyama 1992) or of massive circumstellar discs (e.g. Bonnell 1994; Bate \& Bonnell 1994; Whitworth et al.\ 1995; Burkert, Bate \& Bodenheimer 1997; Hennebelle et al.\ 2004).  Star-disc encounters play a role in truncating discs (Section \ref{ppdiscs}), but they do not play a significant role in forming binary and multiple systems from unbound objects (c.f.\ Clarke \& Pringle 1991a).  Two star-disc encounters resulted in the formation of multiple systems in Calculation 1.  In Calculation 2, there is no obvious example of a multiple system being formed via a star-disc encounter.  However, in both calculations, it is important to note that, although star-disc encounters do not usually form simple bound systems directly, they do result in dissipation which aids in the formation of the small-$N$ bound groups that later dissolve and produce binary and multiple systems.  Thus, dissipative encounters play an important role in star formation (c.f.\ Larson 2002), though not through the simple picture of star-disc capture.

\subsubsection{Multiplicity}

\begin{figure}
\centerline{\psfig{figure=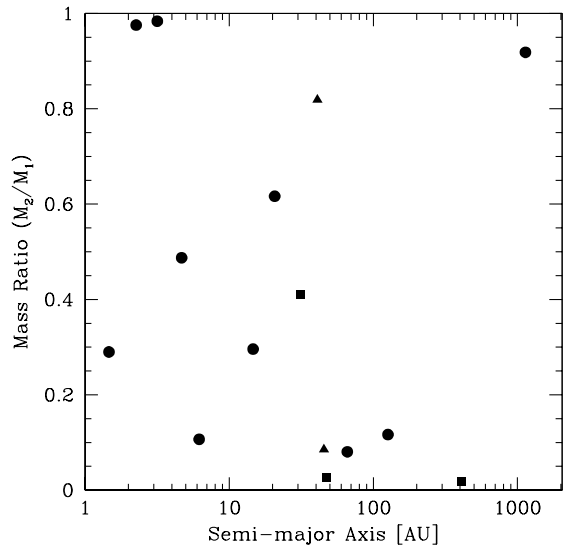,width=8.0truecm}}
\caption{\label{binaryq} Mass ratios versus semi-major axes of the binary, triple and
quadruple systems that exist at the end of Calculation 2 (see also Table 3).  Binaries are plotted 
with circles, triples with triangles and quadruple systems with squares.  This figure should be
compared with Figure 12 of BBB2003 for the equivalent results
from Calculation 1.  Whereas in Calculation 1 there were no wide binaries (separations 
$>10$ AU) and no binaries with mass ratios $M_2/M_1 \la 0.3$, Calculation 2 produces
five wide binaries and three binaries with mass ratios $M_2/M_1<0.2$.}
\end{figure}

When Calculation 2 was stopped, there were 7 distinct multiple systems with semi-major axes $\lsim 1000$ AU.  Their properties are displayed in Table 3 and in Figure \ref{binaryq}.  All but one of these systems originated in the main dense core.  There are ten binaries, three of which (15,33; 20,32; and 14,2) were ejected from the core.  Binary 69,58 is still very weakly bound to the main dense core, but is very isolated at the end of the calculation and so is unlikely to evolve further.  The system (((7,10),26),41) consisting of 4 objects (a close stellar binary and two wider brown-dwarf companions) is also weakly bound to the main dense core but at a very large distance.  The main dense core contains a large bound system of 23 objects, including 3 binaries with separations of 15 AU or less.  Dense core 2 contains a system of 8 objects, including a hierarchical quadruple system.  Cores 3 and 4 each contain a single star.  The remaining 34 objects are either completely unbound or are very weakly bound to the main dense core but have been ejected to very large distances.  

The multiple systems present at the end of Calculation 1 were discussed in detail in BBB2003 and Bate et al.\ (2002b).  Our goal here is to compare and contrast the multiple systems obtained from the two calculations.  BBB2003 had three main conclusions from Calculation 1 regarding multiplicity.  First, when the calculation was stopped, the companion star frequency was high, in broad agreement with observations of star-forming regions.  Second, a large frequency of close binary systems (separations $\lsim 10$ AU) were formed through a combination of dynamical encounters between objects, accretion onto existing multiple systems, and the interaction of multiple systems with circumbinary or circumtriple discs.  This conclusion and the properties of close binary systems that resulted from these formation mechanisms were the topic of Bate et al.\ (2002b).  Due to these formation mechanisms, the close binaries had a preference for equal masses and the frequency of close binaries increased with primary mass.  Third, Calculation 1 produced no wide or low mass-ratio binary systems (separations greater than 10 AU, mass ratios $M_2/M_1<0.25$).  The only wide or low-mass companions were members of higher-order systems (triples, quadruples and higher).  This lack of wide and low mass-ratio binaries also occurs in simulations of $N=5$ clusters embedded in molecular cloud cores (Delgado-Donate et al.\ 2003) and smaller-scale turbulent star-formation calculations (Delgado-Donate et al.\ 2004b).  It seems to be a general result from $N$-body dynamics in small-N clusters and is potentially a serious difficulty because observations suggest there are many unequal-mass wide binaries (e.g.\ Duquennoy \& Mayor 1991).  One possible explanation (Delgado-Donate et al.\ 2004b) is that the frequency of triple and higher-order systems is underestimated observationally and what appear to be wide binaries are in fact multiple systems.

As in the first calculation, Calculation 2 produces a high companion star fraction
\begin{equation}
CSF=\frac{B+2T+3Q+...}{S+B+T+Q+...}
\end{equation}
where $S$ is the number of single stars, $B$ is the number of binaries, $T$ is the number of triples, etc.  The 36 singles, 4 binaries, one quadruple, one octet, and one system of 23 objects give a companion star frequency of $36/43 = 84$ percent.  Alternately, the number of companions divided by the total number of objects is $36/79=46$ percent.  These percentages are higher than in Calculation 1.  Although the systems with more than two components will continue to evolve and will almost certainly eject members, especially the large system of 23 objects, it is plausible that the final companion star frequency will be high, as required by observations of star-forming regions (Ghez, Neugebauer \& Matthews 1993; Leinert et al.\ 1993; Richichi et al.\ 1994; Simon et al.\ 1995; Ghez et al.\ 1997; Duch\^ene 1999).  Disregarding the system of 23 objects entirely still gives a companion star frequency of $14/42 = 33$ percent and the number of companions divided by the total number of objects to be $14/56=25$ percent.

As with Calculation 1, Calculation 2 produces a realistic frequency of close binaries (separations $<10$ AU). Even if all wider systems break up, the resulting frequency of close binaries would be $5/74\approx7$ percent.  This is about a factor of two lower than Calculation 1 and about a factor of three lower than the observational value of $\approx 20$ percent (Duquennoy \& Mayor 1991).  However, Duquennoy \& Mayor were not sensitive to brown dwarfs.  If only stars are considered, the frequency of close binaries becomes $4/15\approx 27$ percent (for Calculation 1, the frequency is almost identical at $5/18\approx 28$ percent).  As in Calculation 1, there is a preference for close binaries to have equal masses (two have mass ratios of $M_2/M_1=0.98$ and only one has a mass ratio lower than $0.25$), and the frequency of close binaries is higher for more massive primaries -- 6 of the 8 most massive stars are members of close binaries, while there is only one binary brown dwarf with separation $<10$ AU out of 46 definite brown dwarfs.  These preferences result from the formation mechanisms of close systems as discussed by Bate et al.\ \shortcite{BatBonBro2002b}.

One of the most interesting results from Calculation 2 is that it produces 5 binaries with separations greater than 10 AU (Figure  \ref{binaryq} and Table 3).  Moreover, three of them have been ejected and shouldn't evolve further and two of these three have low mass ratios ($M_2/M_1 \lsim 0.1$).  As mentioned above, wide binaries and low mass-ratio binaries are absent in Calculation 1 and rare in other numerical simulations of small-N clusters.  The three with the largest separations (66, 126 and 1136 AU) were formed from objects that happened to be ejected from the main dense core at roughly the same time, in approximately the same direction, and at similar speeds.  Thus, as they left the main dense core, they were bound to each other.  As might be expected, they all have large eccentricities (ranging from 0.52 to 0.98).  The reason Calculation 2 produces these wide systems, while Calculation 1 does not, appears to be due to the larger number of objects formed in the main dense core of Calculation 2 and the fact that as this small cluster collapses to its minimum size, a large number of objects are ejected almost simultaneously.  In Calculation 1, the smaller number of objects meant that the intervals between ejection events were longer.  Whether or not such bursts of ejected objects can explain the observed number of wide binary systems is unclear.  However, it may be plausible if a significant fraction of stars form in groups of $\ga 40$ objects.

\begin{table*}
\begin{tabular}{lll}\hline
Disc Radius & Encircled Objects & Comments \\
\multicolumn{1}{c}{AU}        &         &          \\\hline
260         & (4,3),8 & Circum-multiple disc (Figure \ref{core1}, $t=1.40 t_{\rm ff}$, right) \\ 
220         & (19,45),(27,49) & Circumquadruple disc (Figure \ref{core2}, $t=1.40 t_{\rm ff}$) \\ 
200         & 30 & Disc around single star \\ 
110         & (7,10),26       & Circumtriple disc (Figure \ref{core1}, $t=1.40 t_{\rm ff}$, left) \\ 
100         & 67    & Disc around single star\\ 
80         & 76    &  Substellar object formed in isolation near end of calculation, would probably become a star \\ 
 30         & 65    & Substellar object formed near end of calculation, still accreting\\ 
 30         & 66    & Substellar object formed near end of calculation, still accreting\\ 
 20         & 73    & Substellar object formed near end of calculation, still accreting\\ 
 \hline
\end{tabular}
\caption{\label{tablediscs} The discs around objects that exist around objects when Calculation 2 is stopped. Discs with radii $\lsim 10$ AU are not resolved.  Unlike Calculation 1, in Calculation 2 no objects are ejected with resolved discs. This table should be
compared with Table 4 of BBB2003 for the equivalent results
from Calculation 1.}
\end{table*}

\subsubsection{Brown-dwarf companions to stars and brown dwarfs}

Calculation 1 produced one binary brown dwarf system out of $\approx 20$ brown dwarfs, implying a frequency of binary brown dwarfs of $\sim 5$ percent (with a large uncertainty).  Furthermore, although it was a close binary, it was still part of an unstable multiple system and was still accreting.  Thus, its long term survival was not certain.  This low frequency of binary brown dwarfs is due to accretion and dynamical interactions.  The binary brown dwarf must avoid accreting too much gas or it will become a stellar binary.  It may stop accreting if it is ejected from the dense molecular gas in which it formed by dynamical encounters (Section 3.6).  However, to produce a {\it binary} brown dwarf, the binary must be ejected as a whole.  To reach escape velocity from the molecular cloud requires a close dynamical interaction.  Thus, rather than be ejected, a binary brown dwarf is likely to be broken up or to have one of its components replaced by a star in an exchange interaction.  The result of this formation mechanism is that wide binary brown dwarfs are very unlikely, and even close systems are likely to be rare since they must be ejected but not undergo an exchange interaction (see also Reipurth \& Clarke 2001; Bate et al.\ 2002a).

However, Calculation 2 shows that there is another mechanism for forming binary brown dwarfs, and these systems tend to be wide.  Three binary brown dwarf (BBD) systems and two additional binaries in which a brown dwarf orbits a very-low-mass (VLM) star ($M<0.09$ M$_\odot$) are formed in Calculation 2.  The properties of these binaries are given in Table \ref{table3}.  One of the BBDs has been ejected (15,33) and will not evolve further.  One of the VLM star/brown dwarf systems (20,32) has also been ejected.  Interestingly, these systems are the wide 66-AU and 126-AU systems mentioned in the previous section. Each formed when two objects were ejected simultaneously in similar directions and, thus, were bound to each other.  The two other BBDs (60,59 and 69,58) are still formally bound to the cluster in the main dense core, however, when the calculation is stopped both are at very large distances.  BBD (60,59) consists of two 9 Jupiter-mass objects and is a very close system (2 AU) so is likely to survive in the long term.  The survival of BBD (69,58) as a binary brown dwarf is less certain as it is wider (21 AU) and both components are still accreting.  The final VLM star/brown dwarf binary (6,54) is close (15 AU), very weakly bound to the main dense core, and both objects are accreting. 

In summary, Calculation 2 gives the overall fraction of VLM star/brown dwarf and BBD systems to be 5 out of 60 systems with component masses less than 0.09 M$_\odot$, i.e. $\approx 8$ percent.  This is somewhat higher than Calculation 1 and has the advantage that at least two of the systems will not evolve further.  The observed frequency of very-low-mass and brown dwarf binaries is $\approx 15$ percent (Reid et al.\ 2001; Close et al.\ 2002, 2003; Bouy et al.\ 2003; Burgasser et al.\ 2003; Gizis et al.\ 2003; Mart{\'{\i}}n et al.\ 2003).  Thus, the calculations under-produce binary brown dwarfs by roughly a factor of two.  However, we are still limited by poor statistics and further calculations are required.

On the observational side, it is very important to determine more completely the period distribution of BBDs.  Until recently, all of the known binary brown dwarfs systems were close (separations $< 15$ AU), consistent with the ejection hypothesis for brown dwarf formation.  However, only one spectroscopic BBD system is currently known \cite{BasMar1999}.  Furthermore, Luhman \shortcite{Luhman2004a} has recently reported the discovery of a BBD candidate with a separation of $\approx 200$ AU.  Although this system may be explained by the mechanism for wide BBD formation described above, an accurate determination of the frequency of very close and wide systems is essential for constraining future models.

Along with a BBD, Calculation 1 one produced one binary system consisting of a star (0.13 M$_\odot$) and a brown dwarf (0.04 M$_\odot$).  The system had a separation of 7 AU and was part of an unstable septuple system.  Both objects were still accreting.  Calculation 2 does not produce any such systems, but there are two stars with VLM stellar companions (binaries 27,49 and 19,45 in Table \ref{table3}).  Both of these binaries have separations less than 10 AU.  In fact, these two binaries form a hierarchical quadruple system that is at the centre of the group of stars in the second dense core when the calculation is stopped.  All four objects are still accreting, so whether or not the two VLM stars will remain as low-mass companions is in doubt.  In any case, brown dwarf companions in close orbits around stars with masses $\gsim 0.1$ M$_\odot$ seem to be rare.  There are two main reasons for this.  First, in order for the primary to become a star it must have accreted a lot of gas.  Any companion would likely become a star also because the long term effect of accretion onto a binary is generally to drive the system to equal mass components (Bate 2000; Whitworth et al.\ 1995).  Second, while the primary is accreting to its stellar mass it is embedded in the molecular cloud in which it formed and is likely to undergo dynamical interactions with other objects (see Section \ref{ppdiscs}).  If these interactions involve exchanges, any low-mass companion will likely be replaced by an object with a higher mass.  

The low frequency of brown dwarfs in close orbits around stars is in agreement with observations.  Doppler searches for planets orbiting solar-type stars find a very low frequency of brown dwarf companions in tight orbits, the so called brown dwarf desert \cite{MarBut2000}.  Wide star/brown dwarf systems are seen observationally \cite{Gizisetal2001}.  Although many of the stars in Calculations 1 and 2 have wide brown dwarf companions, because these systems are still dynamically evolving when the calculations are stopped we cannot compare our calculations with these observations.  However, the small-scale turbulent star-formation simulations of Delgado-Donate et al.\ (2004b), which were evolved until the systems reached dynamical stability, do predict that many close stellar binary systems should have wide brown dwarf companions.

\begin{figure}
\centerline{\psfig{figure=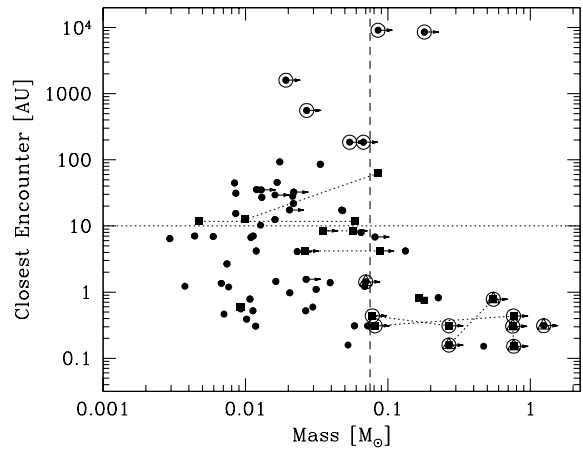,width=8.0truecm}}
\caption{\label{truncate}  The closest encounter distance of each 
star or brown dwarf during Calculation 2 versus the object's final mass.  This figure should be
compared with Figure 14 of BBB2003 for the equivalent results
from Calculation 1.  Objects that 
are still accreting significantly at the end of the calculation
are denoted with arrows indicating that they are still evolving and
that their masses are lower limits.  Objects that
have resolved discs at the end of the simulation are circled.
Discs smaller than $\approx 10$ AU (horizontal dotted line) cannot 
be resolved by the simulation.  Objects that have had close 
encounters may still have resolved discs due to subsequent accretion
from the cloud.  Note that there are only 9 resolved discs 
at the end of the simulation, but many surround binary and 
higher-order multiple systems (hence the 16 circles in the figure).
Close binaries (semi-major axes 
$<10$ AU) are plotted with the two components connected by dotted 
lines and squares are used as opposed to circles.  
Components of triple systems whose orbits have semi-major 
axes $10<a<100$ AU are denoted by triangles.  All but one of the 
close binaries is surrounded by a resolved disc.  
Encounter distances less than 4 AU are upper limits since the point 
mass potential is softened within this radius.  The vertical 
dashed line marks the star/brown dwarf boundary.  The four
brown dwarfs in the top left corner of the
figure that are still accreting formed shortly before the calculation
was stopped are thus still evolving 
rapidly.  They may not end up as brown dwarfs or with resolved
discs.  There are no brown dwarfs that have resolved discs and have 
finished accreting.} 
\end{figure}

\subsection{Protoplanetary discs}
\label{ppdiscs}

The calculations resolve gaseous discs with radii $\ga 10$ AU around the young stars and brown dwarfs.  Discs with typical radii of $\sim 50$ AU form around many of the objects due to the infall of gas with high specific angular momentum.  However, one of the surprises from Calculation 1 was that most of these discs were severely truncated in subsequent dynamical interactions.  By the end of the calculation, most of the discs were too small to have formed our solar system (BBB2003).  Only two objects, one star and one brown dwarf, were ejected from the cloud with resolved discs (radii of 50 AU and 60 AU, respectively).  Nine other discs ranging in radius from 20-200 AU existed around objects when the calculation was stopped, but these objects were still members of unstable multiple systems.

In Calculation 2, because of the higher stellar densities reached in the dense molecular cloud cores, the situation is even worse for the survival of large discs.  With 79 objects, there are only 9 resolved discs (compared to 11 resolved discs among 50 objects for Calculation 1).  In Figure \ref{truncate}, we plot the closest encounter distance for each object during the calculation as a function of its final mass.  Table \ref{tablediscs} lists the properties of the 9 resolved discs at the end of the calculation.  All but three stars have had encounters closer than 10 AU.  The two stars with greatest encounter distances ($\approx 10^4$ AU) are those two stars that formed on their own in dense cores 3 and 4.  These two stars (objects 30 and 67) have disc radii of $\approx 200$ and $\approx 100$ AU, respectively.  The other stars with resolved discs are all members of multiple systems surrounded by large discs.  Although they have had very close encounters, subsequent infalling gas has build up circumtriple and circumquadruple discs around them.  Most of the substellar objects have also had encounters closer than 10 AU.  Four of them are surrounded by resolved discs at the end of the calculation (Table \ref{tablediscs}) with radii ranging from approximately 20 to 80 AU.  However, none of the stars or brown dwarfs ejected during Calculation 2 were surrounded by resolved discs.

Calculation 1 modelled a dense star-forming region, so Calculation 2 is even more extreme.  The primary motivation for Calculation 2 is to investigate the dependence of the IMF on the mean thermal Jeans mass in molecular clouds.  Such high-density initial conditions are not meant to be representative of local star formation.  Calculation 2 does, however, confirm that the prolific disc truncation seen in Calculation 1 is a general feature of such simulations and that, as expected, the resulting size distribution of discs moves to smaller radii discs with increasing stellar density.

\section{Discussion}

\subsection{The observed abundance of brown dwarfs}

The above results clearly show that, all other things being equal, there should be a greater proportion of brown dwarfs in star-forming clouds with lower mean thermal Jeans masses.  We find that reducing the mean thermal Jeans mass by a factor of three increases the fraction of brown dwarfs by about a factor of 1.5 (from $\approx 50$ to $\approx 75$ percent).  Thus we would expect about a factor of two difference in the frequency of brown dwarfs for a change of one order of magnitude in the Jeans mass. Are there any observations to support this hypothesis?

Brice{\~ n}o et al. \shortcite{Bricenoetal2002}, compared the IMFs in Taurus and the Orion Trapezium cluster.  They found a factor of two fewer brown dwarfs in Taurus.  New results on the Trapezium cluster \cite{SleHilCar2004} have decreased this discrepancy to a factor of 1.5, but Taurus still appears to be deficient in brown dwarfs \cite{Luhman2004b}.  Brice{\~ n}o et al. \shortcite{Bricenoetal2002}, proposed that this difference may be due to the different mean densities (and hence mean Jeans masses) of the regions.  Taurus is a low-mass low-density star-forming region.  It is difficult to define a mean thermal Jeans mass in the region because the cloud is so patchy.  However, the gas mass of the Taurus dark cloud is measured to be $\approx 10^4$ M$_\odot$ in a radius of $\approx 10$ pc (Baud \& Wouterloot 1980; Cernicharo, Bachiller \& Duvert 1985).  This gives a mean thermal Jeans mass of $\approx 30$ M$_\odot$, though the local Jeans mass in some of the dense cores is more than an order of magnitude lower than this.  The Trapezium cluster is a high-mass high-density star-forming region.  It contains approximately $5\times 10^3$ M$_\odot$ of stars and gas (Hillenbrand \& Hartmann 1998) within a radius $\approx 2$ pc, giving a mean thermal Jeans mass of $\approx 3$ M$_\odot$ if it is assumed that the temperature before star formation began was $\approx 10$ K.  Thus, the difference in the mean thermal Jeans mass between the two regions is probably about an order of magnitude.  A factor of $\approx 1.5-2$ fewer brown dwarfs in Taurus is therefore in agreement with both the direction and magnitude of the above prediction.

Calculation 1 had a mean thermal Jeans mass of 1 M$_\odot$, similar to the progenitor cloud of the Trapezium (perhaps slightly lower).  Both calculations have mean Jeans masses significantly lower than Taurus.  As discussed in Section 2.3, this is purely for computational reasons since with current computational resources we are only able to fully resolve the fragmentation of $\approx 50$ M$_\odot$ of gas but we wish to model clouds containing many thermal Jeans masses.  Thus, the frequencies of brown dwarfs in these calculations should be significantly higher than observed in Taurus (as, in fact, they are; Briceno et al.\ 2002; Luhman 2004b) but Calculation 1 should give a frequency similar to that of the Trapezium cluster.  Calculation 1 produces roughly equal numbers of stars and brown dwarfs, but we emphasise that the calculation resolves objects with masses down to the opacity limit for fragmentation of a few Jupiter masses.  Current IMF determinations in the Trapezium cluster are only complete down to $\approx 0.02$ M$_\odot$.  These surveys find that about 20\% of the objects are brown dwarfs (Hillenbrand \& Carpenter 2000; Slesnick et al.\ 2004).  Taking objects with masses greater than $0.02$ M$_\odot$ in Calculation 1, we find $13/39\approx 33$\% are brown dwarfs.  Given our small number statistics and the fact that the mean Jeans mass in the calculation (1 M$_\odot$) may be slightly lower than in the progenitor cloud of the Trapezium cluster ($\approx 3$ M$_\odot$), these numbers are in reasonable agreement.  The implication is that extending observational surveys in the Trapezium cluster down to the opacity limit for fragmentation may increase the number of brown dwarfs by up to 50 percent.

\subsection{A simple model for the IMF}

\begin{figure*}
\centerline{\psfig{figure=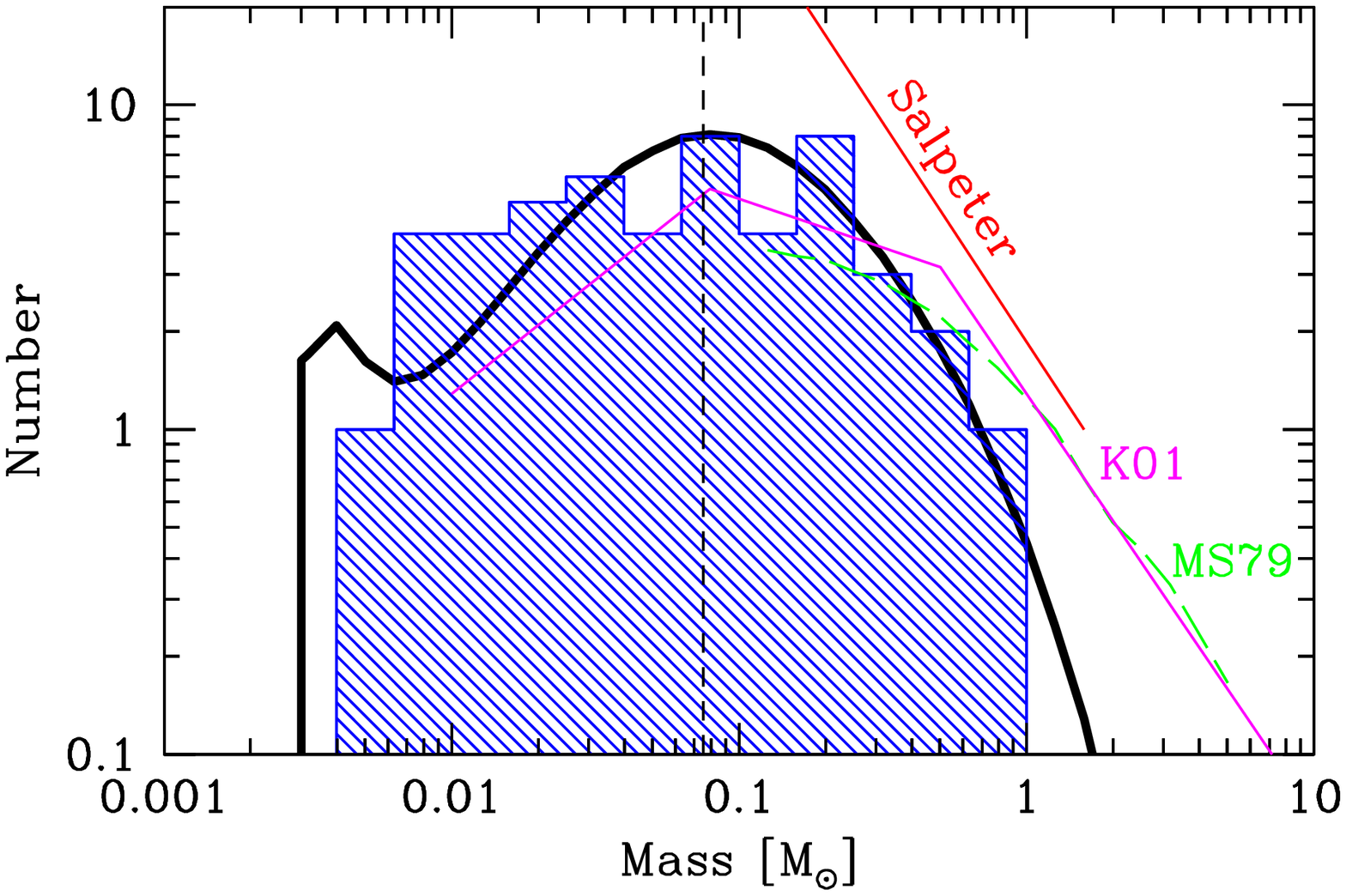,width=9.0truecm}\psfig{figure=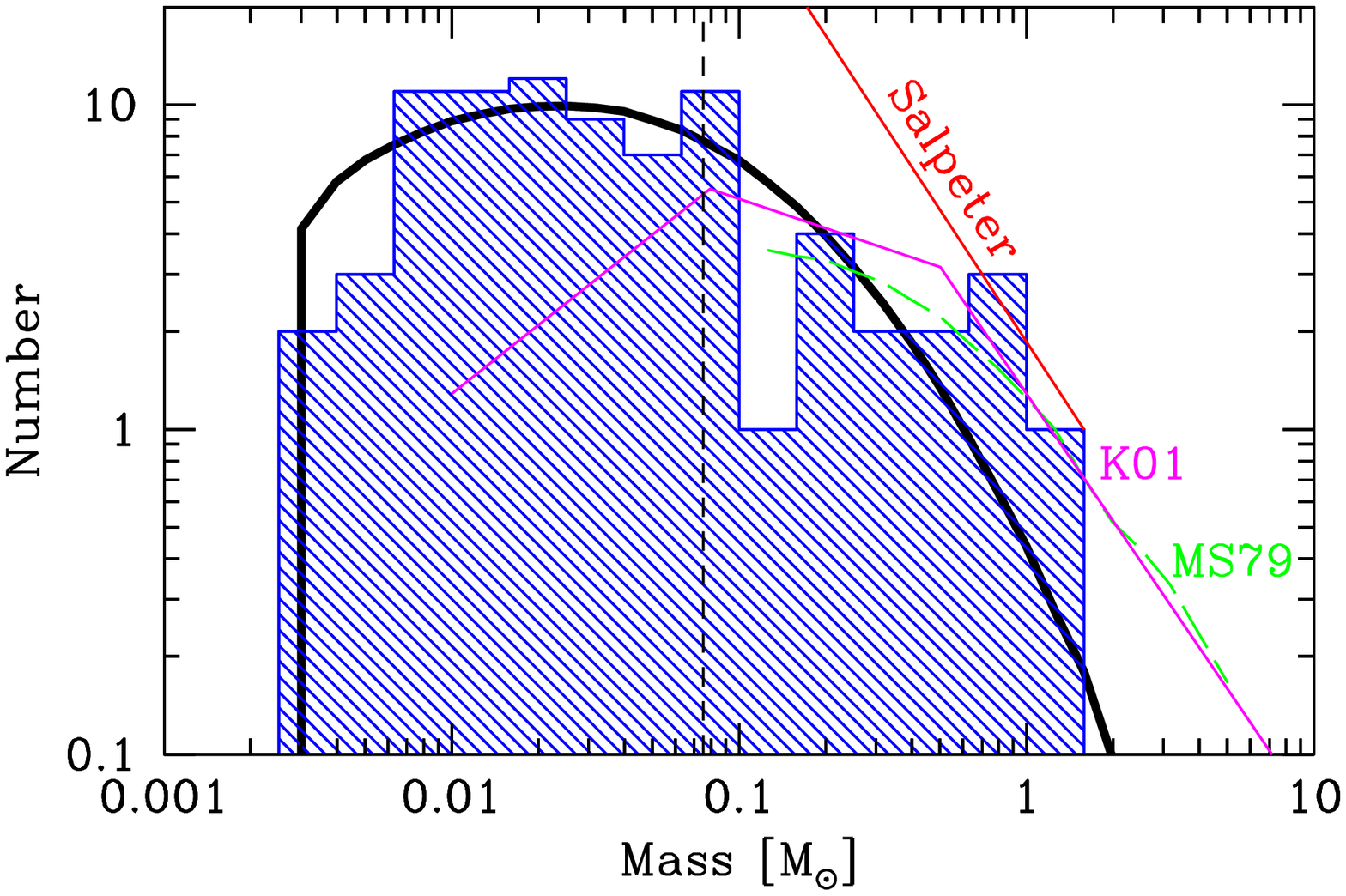,width=9.0truecm}}
\caption{\label{imffit} The initial mass functions produced by the two hydrodynamical calculations (histograms) and their fits using the simple accretion/ejection IMF model (thick solid line).  Statistically, the hydrodynamical and the model IMFs are indistinguishable.  The panel on the right gives the result for the denser cloud that has the lower initial mean thermal Jeans mass.  Also shown are the Salpeter slope (solid straight line), and the Kroupa (2001) (solid broken line) and Miller \& Scalo (1979) (dashed line) mass functions.  The vertical dashed line is the stellar-substellar boundary.} 
\end{figure*}

We have argued that the IMFs produced by the two hydrodynamical calculations discussed in this paper may be understood as originating from a combination of accretion and dynamical ejections which terminate the accretion.  Based on this finding, we develop a very simple model for the origin of the IMF.   We find it reproduces the IMFs obtained in the two calculations very well and, using the values of the parameters obtained from the simulations, produces a near-Salpeter slope at high masses.

Consider a star-forming molecular cloud.  The simple accretion/ejection model for the IMF is as follows.  
\begin{itemize}
\item We assume all objects begin with masses set by the opacity limit for fragmentation ($\approx 3$ M$_{\rm J}$ for the calculations presented here) and then accrete at a fixed rate $\dot{M}$ until they are ejected.
\item We assume the accretion rates of individual objects are drawn from a log-normal distribution with a mean accretion rate (in log-space) given by $\log(\overline{\dot{M}})=\overline{\log(\dot{M})}$ and a dispersion of $\sigma$ dex (i.e. $\log(\dot{M}) = \log(\overline{\dot{M}}) + \sigma G$, where $G$ is a random Gaussian deviate with zero mean and unity variance).
\item The ejection of protostars from an $N$-body system is a stochastic process.  It cannot be solved analytically and must be described in terms of the half-life of the process.  We assume that there is a single parameter, $\tau_{\rm eject}$, that is the characteristic timescale between the formation of an object and its ejection from the cloud.  The probability of an individual object being ejected is then $\exp(-t/\tau_{\rm eject})$ where $t$ is the time elapsed since its formation.  Note that a similar assumption was used by Reipurth \& Clarke \shortcite{ReiCla2001} when they considered the ejection of brown dwarfs from unstable triple systems.
\end{itemize}
Clearly, these are assumptions involve gross simplifications.  The accretion rates of individual objects do vary with time and it is not clear that the dispersion in the time-averaged accretion rates of objects is log-normal.  In particular, we note that objects that end up with a low mass have a larger dispersion in their time-averaged accretion rates than those that accrete over a long period and end up as high-mass objects (Figure \ref{accrates}).  This indicates that the accretion rates are variable on short timescales, but the long-term averages may be less variable.  Also, the timescale for the ejection of an object from the cloud must depend on its local situation.  Despite these objections, over a large number of objects, one might hope that these assumptions are are reasonable description of the behaviour of a typical object (e.g.\ in Figure \ref{accmass}, the mass of an object seems to depend linearly on the time it spends accreting with only a small dispersion).

Assuming that the cloud forms a large number of objects, $N$, and that the time it evolves for is much greater than the characteristic ejection time, $T\gg \tau_{\rm eject}$, then there are essentially only three free parameters in this model.  These are the mean accretion rate times the ejection timescale, $\overline{M}=\overline{\dot{M}}\tau_{\rm eject}$, the dispersion in the time-averaged accretion rates, $\sigma$, and the minimum mass provided by the opacity limit for fragmentation, $M_{\rm min}$.  If $\overline{M}>>M_{\rm min}$, $\overline{M}$ is the characteristic mass of an object.  For the hydrodynamical calculations in this paper, the minimum mass is the same, roughly 3 M$_{\rm J}$.  Thus, there are only two free parameters.  

\begin{table}
\begin{tabular}{lcccc}\hline
Model\hspace{-10pt} & $\overline{\dot{M}}$ & $\sigma$ & $\tau_{\rm eject}$  & $T$\\
       &    M$_\odot$/yr     &  Dex.    & yr  &  yr    \\\hline
1 & $6.17\times 10^{-6}$ & 0.33 & $3.2\times 10^4$ & $6.91\times 10^4$ \\
2 & $7.18\times 10^{-6}$ & 0.50 &  $9.3\times 10^3$ &   $3.67\times 10^4$\\
 \hline
\end{tabular}
\caption{\label{tableimf} The parameters of the simple accretion/ejection IMF models that fit the IMFs from the hydrodynamical calculations (Figure \ref{imffit}).  There are essentially two parameters in the models, the mean accretion rate times the characteristic timescale for ejection $(\overline{\dot{M}}\tau_{\rm eject})$ and the dispersion in the accretion rates $\sigma$.  The time period over which the simulations are run, $T$, has a small effect on the form of the IMF.  For example, in the left-hand panel, the peak in the model IMF at very low masses is because two objects were formed shortly before the calculations were stopped and therefore these two objects do not usually manage to accrete much mass in the model.}
\end{table}

 \subsubsection{Reproduction of the hydrodynamical IMFs}

The hydrodynamical calculations are not followed until all the stars and brown dwarfs have finished accreting (i.e., the IMF is not fully formed).  It is not the case that $T\gg \tau_{\rm eject}$.  This must be taken into account when calculating simple accretion/ejection models for comparison with the results of the hydrodynamical calculations.  To do this, we must evolve the simple models over the same periods of time that the hydrodynamical simulations took to form their stars and brown dwarfs (i.e. from $T=1.04-1.40 t_{\rm ff}$ and $T=0.82-1.40 t_{\rm ff}$ for Calculations 1 and 2, respectively). The times of formation of each of the objects are taken directly from the hydrodynamical simulations (i.e.\ from Figure \ref{sfrate} for Calculation 2 and Figure 7 of BBB2003 for Calculation 1).

We then generate model IMFs for comparison with the results of the two hydrodynamical calculations (Figure \ref{imffit}).  Each model IMF is the average of 30000 random realisations of the simple accretion/ejection model, keeping the values of the input parameters fixed.  The parameter values are given in Table \ref{tableimf}.  
It is important to note that these parameters were {\it not} varied in order to obtain good fits to the hydrodynamical IMFs.  Rather, {\it the values of the parameters were taken directly from the hydrodynamical simulations}.  There is no freedom to vary the parameters in order to obtain a better fit. 
The mean accretion rate of the objects, $\overline{\dot{M}}$, and the dispersion in the accretion rates, $\sigma$, were set equal to the mean (in log-space) of the time-averaged accretion rates and their dispersion from Figure \ref{accrates}.  The characteristic ejection times, $\tau_{\rm eject}$, were set so that the mean numbers of objects ejected from the two sets of 30000 random realisations matched the number of objects ejected during each of the hydrodynamical calculations (26 and 50 for Calculations 1 and 2, respectively).

Figure \ref{imffit} shows that the simple accretion/ejection models match the hydrodynamical IMFs very well.  Kolmogorov-Smirnov tests show that the hydrodynamical IMFs from Calculations 1 and 2 have 92 and 27 percent probabilities of being drawn from the model IMFs, respectively (i.e.\ they are consistent with each other).  Thus, we demonstrate that a simple model of the interplay between accretion and ejection can reproduce the low-mass IMFs produced by the hydrodynamical calculations and give a near-Salpeter slope for high masses ($M\gsim 0.5$ M$_\odot$).

\subsubsection{An analytical form for the accretion/ejection IMF model}
\label{analytical}

In the limit that a cloud forms a large number of objects, $N$, and that the time it evolves for is much greater than the characteristic ejection time, $T\gg \tau_{\rm eject}$ (i.e.\ all objects have finished accreting and the star formation is complete, neither of which is true for the IMFs obtained from the hydrodynamical calculations), the simple accretion/ejection IMF model can be formulated semi-analytically rather than requiring Monte-Carlo simulation.

The probability distribution of the individual accretion rates for the stars and brown dwarfs is assumed to be log-normal
\begin{equation}
\label{probaccrate}
p(\dot{M}) = \frac{1}{\sqrt{2\pi}\sigma \dot{M}}\exp{\left(\frac{-\left(\log{\dot{M}} - \log{\overline{\dot{M}}}\right)^2}{2\sigma^2}\right)}.
\end{equation}
The final mass of an object is
\begin{equation}
\label{finalmass}
M = M_{\rm min} + \dot{M} t,
\end{equation}
where $t$ is the time between the formation of the object and its accretion being terminated.
We now require the probability distribution of the masses of the objects $f(M)$, at time $t$ (assuming all objects accrete indefinitely).  This is obtained noting that
\begin{equation}
p(\dot{M}){\rm d}{\dot{M}}=f(M){\rm d}M,
\end{equation}
so that
\begin{equation}
f(M) = p(\dot{M})\frac{{\rm d}{\dot{M}}}{{\rm d}M} = \frac{p(\dot{M})}{t}.
\end{equation}
Rearranging equation \ref{finalmass} for the accretion rate and substituting this into equation \ref{probaccrate} gives 
\begin{equation}
f(M,t) = \frac{1}{\sqrt{2\pi}\sigma (M-M_{\rm min})} \exp{\left(\frac{-\left(\log{\left(\frac{M-M_{\rm min}}{t}\right)} - \log{\overline{\dot{M}}}\right)^2}{2\sigma^2}\right)},
\end{equation}
for $M>M_{\rm min}$.  Finally, we need to take account of the fact that objects are ejected, terminating their accretion stochastically.  The probability an object is ejected at time $t$ is
\begin{equation}
e(t) = \frac{1}{\tau_{\rm eject}}\exp{\left(-\frac{t}{\tau_{\rm eject}}\right)}.
\end{equation}
Thus, the final mass function is
\begin{equation}
f(M) = \int_{0}^{\infty} f(M,t)e(t) {\rm d}t.
\end{equation}
This cannot be integrated analytically, but it is trivial to integrate numerically.  Examples of the resulting mass function are shown in Figure \ref{imfmodels} and discussed below.

We note that this model has some similarities with the IMF models of Myers \shortcite{Myers2000} and Basu \& Jones \shortcite{BasJon2004}.  They also propose that the accretion of individual objects is terminated stochastically.  However, there are several differences between their models and ours.  Myers proposes that the masses of cores grow with time and this accretion is terminated when the core is triggered to collapse.  Basu \& Jones discuss accretion on to protostars, terminated by {\it any} stochastic process (e.g.\ dynamical ejections).  Furthermore, they both propose that the initial masses of the objects are drawn from a log-normal distribution and that their accretion rates increase with time in proportion to their masses.  We postulate that all objects begin with the {\it same} mass (due to the opacity limit) but have accretion rates drawn from a log-normal distribution that are {\it held constant} until their accretion is terminated.  These differences result in different forms for the IMF.  The model of Basu \& Jones gives a log-normal shape at low-masses, switching to a power-law at high-masses.  Our model has a cut-off in the IMF at low-masses and does not achieve a pure power-law slope at high-masses.

\subsubsection{Variations in the IMF with environment}
\label{variation}

\begin{figure}
\centerline{\psfig{figure=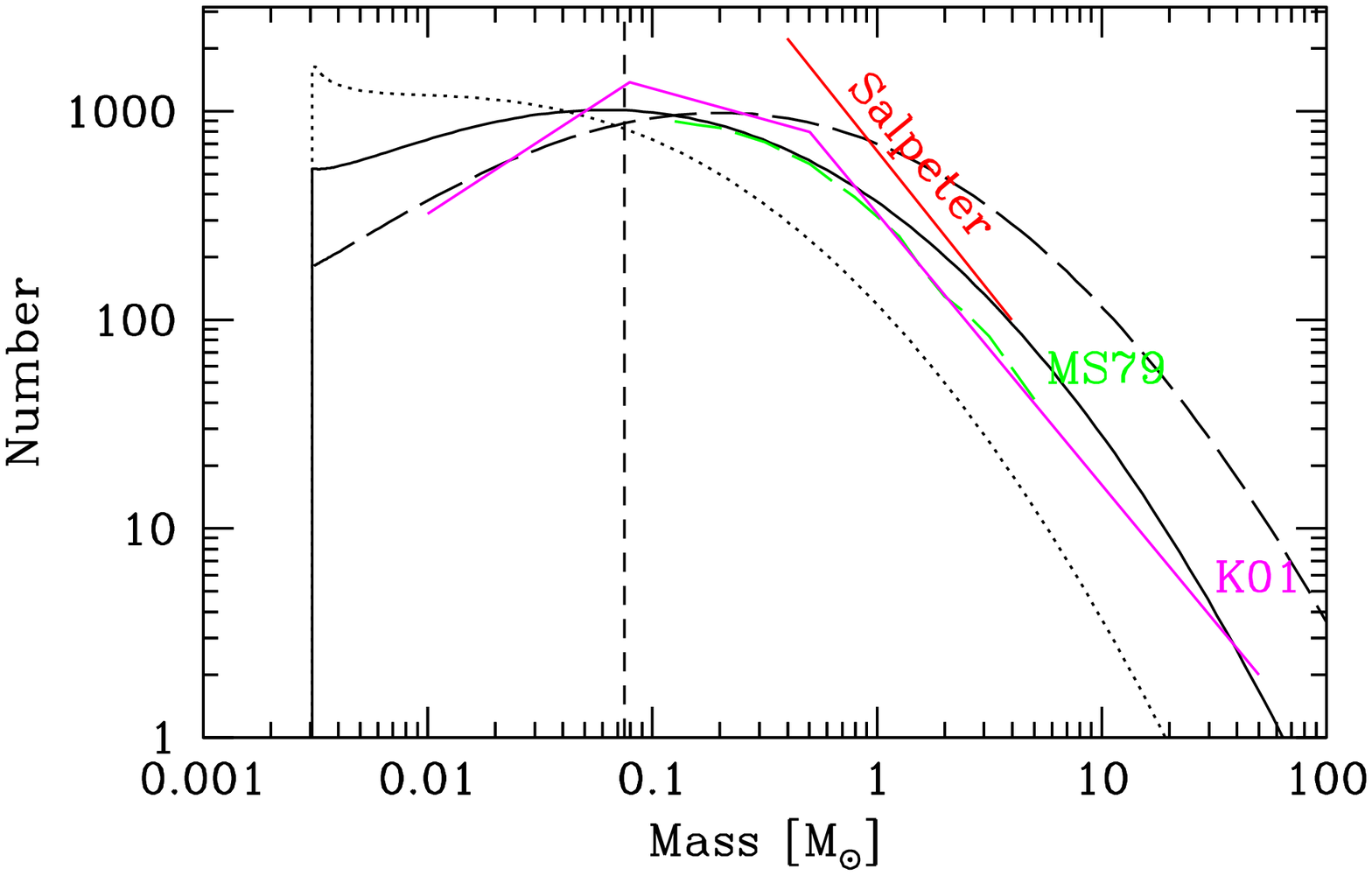,width=8.0truecm}}
\centerline{\psfig{figure=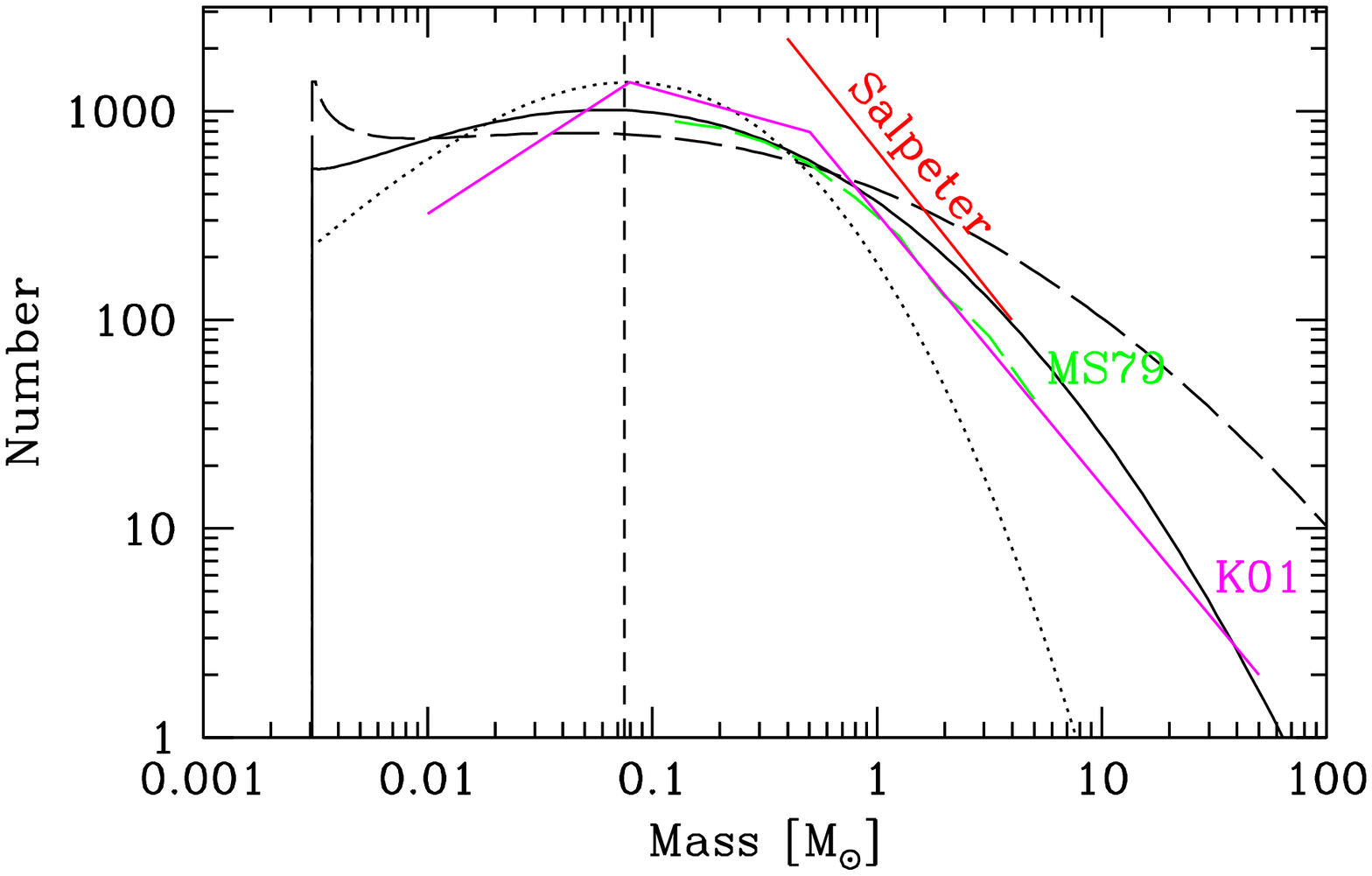,width=8.0truecm}}
\centerline{\psfig{figure=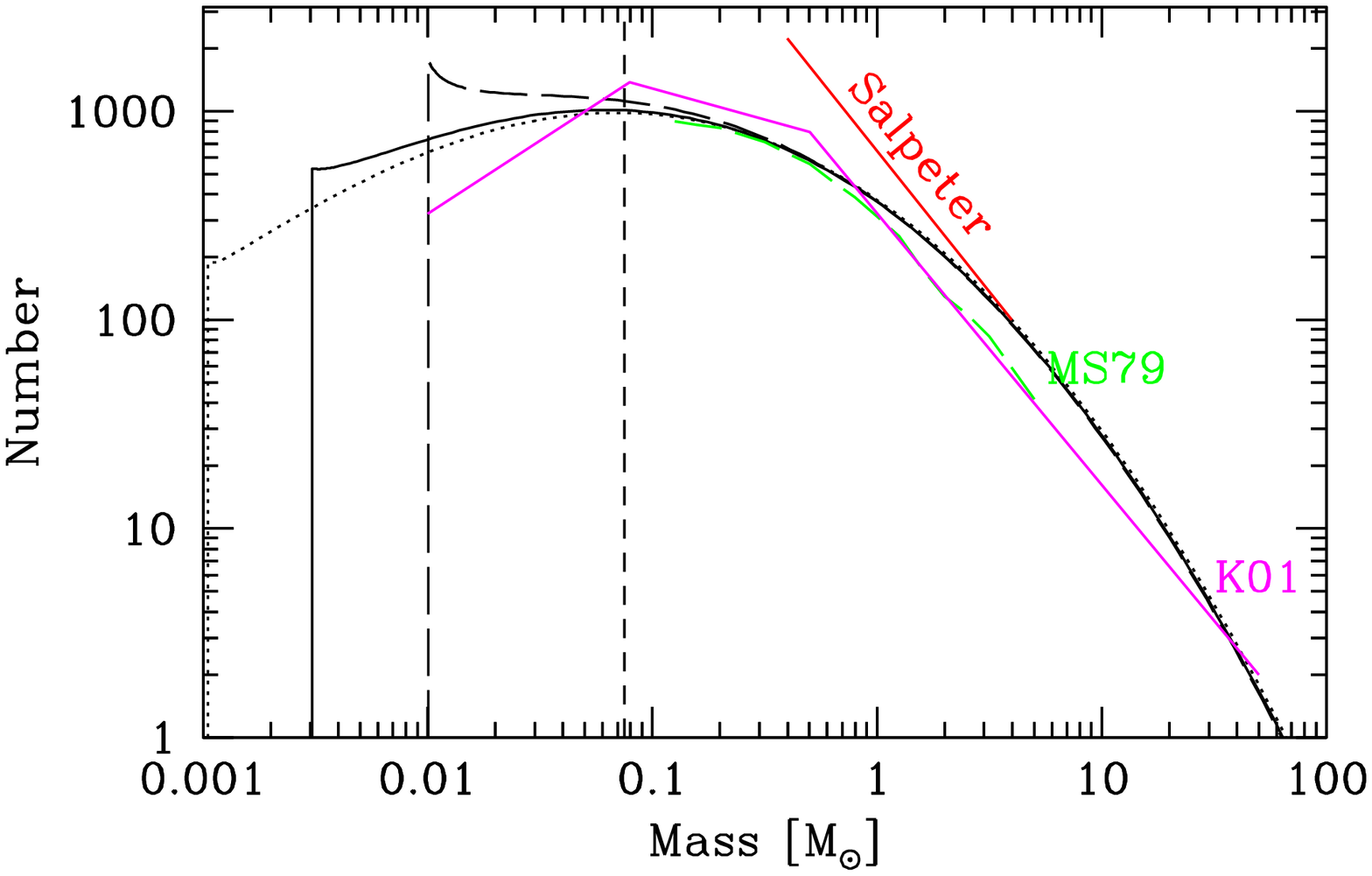,width=8.0truecm}}
\caption{\label{imfmodels} The initial mass functions produced by the simple accretion/ejection IMF model.  The standard case has $\overline{M}=0.1$ M$_\odot$, $\sigma=0.7$, and $M_{\rm min}=0.003$ M$_\odot$ (solid line).  In the top panel, we vary $\overline{M}$ giving results for 0.03  (dotted line) and 0.3  (long-dashed line) M$_\odot$.  In the middle panel, we vary $\sigma$ giving results for 0.4 (dotted line) and 1.0 (long-dashed line).  In the bottom panel, we vary $M_{\rm min}$ giving results for 0.001 (dotted line) and 0.01 (long-dashed line) M$_\odot$.   Also shown are the Salpeter slope, and the Kroupa (2001) and Miller \& Scalo (1979) mass functions.  The vertical dashed line is the stellar-substellar boundary.} 
\end{figure}

In Figure \ref{imfmodels}, we show how the IMF produced by the simple accretion/ejection model varies as a function of the three free parameters.  The parameter $\overline{M}$, which is the mean accretion rate times the characteristic timescale for ejections in the star-forming region, gives the approximate location of the peak in the IMF (i.e.\ the characteristic mass).  The dispersion in the accretion rates $\sigma$ sets the breadth of the IMF and, thus, alters the slopes at the low-mass and high-mass ends.  Dispersions of $\sigma\approx 0.7$ dex give a Salpeter-type slope at high masses.  Finally, the minimum mass, $M_{\rm min}$, sets the low-mass cut-off of the IMF, but otherwise has little effect on the form of the IMF.

To allow the simple accretion/ejection model to be tested by observations, we need to know how these three parameters scale with the physical properties of star-forming regions.  Note that the part of the IMF that is most sensitive to variations in the input parameters is the low-mass end of the IMF.  The effect of altering any of the parameters is to change the turn-over mass, the substellar slope, and/or the low-mass cut-off.  Only a change in the value of the dispersion parameter, $\sigma$, alters the slope of the high-mass IMF.  

As yet, the minimum mass cut-off has not been detected observationally.  For population I and II stars, it is expected to scale with metallicity as $Z^{-1/7}$ (Low \& Lynden-Bell 1976), a very weak dependence that will be difficult to confirm through observations.  

It is not immediately clear how the dispersion in the accretion rates of individual objects should depend on environment.  However, we do find from the two hydrodynamical calculations that the dispersion is greater for the denser cloud (see Table 5).  This implies that the dispersion in accretion rates may be a function of the cloud density, although it would be desirable to test this with further calculations.  However, this possible dependence of the IMF on environment should be able to be tested by observations already by looking to see if there is any indication that the slope of the high-mass IMF is shallower for denser star-forming regions.

The remaining parameter is the characteristic mass, $\overline{M}$.  This depends on the mean accretion rate and the characteristic timescale for ejections.  On dimensional grounds, the former may be expected to scale roughly as $c_{\rm s}^3/G$.  Taking the isothermal sound speed from the calculations, this gives an expected accretion rate of $1.5\times 10^{-6}$ M$_\odot$/yr (the sound speed is the same in each of the hydrodynamical calculations).  The mean accretion rates (in log-space) of the objects are almost identical for the two hydrodynamical calculations (as expected) with values of $\approx 6-7\times 10^{-6}$ M$_\odot$/yr (Table 5).  This is somewhat higher than $c_{\rm s}^3/G$, but this is consistent with previous calculations that show that the accretion rates of objects formed from the collapse of non-singular isothermal spheres are typically somewhat greater than $c_{\rm s}^3/G$ (e.g.\ Foster \& Chevalier 1993).  The ejection timescale of a small $N$-body system should scale with the crossing timescale of the system which in turn scales roughly as $1/\sqrt{G\rho}$, where $\rho$ is the mass density of the system.  
The pattern of the initial turbulence in Calculations 1 and 2 is identical.  The dense cores form due to converging flows in this initial velocity field and, thus, are similar in locations and masses for the two calculations (compare Figure 2 and Table 1 of BBB2003 with Figure 1 and Table 2 of this paper).  However, since the overall cloud is smaller and denser in Calculation 2, the dense cores and resulting stellar groups are smaller and denser by the same factor (i.e.\ the density of the stellar groups is proportional to the density of the progenitor cloud).  Thus, the characteristic timescale for ejections $\tau_{\rm eject}$ should scale inversely with the square root of the initial density of the cloud.  Indeed, the value of $\tau_{\rm eject}$ for Calculation 2 is almost exactly a factor of three smaller than that for Calculation 1 (see Table 5).  Thus, we expect that the characteristic mass $\overline{M}$, the product of the mean accretion rate and the characteristic timescale for ejections should scale as $c_{\rm s}^3/\sqrt{G^3 \rho}$.  Neglecting constants of order unity, {\it this is the definition of the mean thermal Jeans mass of the progenitor clouds}.  

The median masses of the stars/brown dwarfs from the two hydrodynamical calculations follow this scaling almost exactly.  The median mass is a factor of 3.04 lower in Calculation 2 compared with Calculation 1 whereas the mean thermal Jeans mass is exactly a factor of 3 lower.  In the simple accretion/ejection IMF models, the characteristic mass essentially gives the location of the peak in the mass function or the median mass.  Although we argue above that this mass should scale with the mean thermal Jeans mass of the cloud and the hydrodynamical calculations support this, we still need to determine the constant of proportionality between the median mass and the mean thermal Jeans mass.  Since the hydrodynamical calculations have mean thermal Jeans masses of 1 and 1/3 M$_\odot$ and median object masses of 0.070 and 0.023 M$_\odot$, we conclude that the peak of the IMF (in ${\rm d}N/{\rm d}\log M$) occurs at $\approx 1/14$ of the mean thermal Jeans mass of the progenitor cloud.

The mean object masses from the hydrodynamical calculations only differ by 17\% rather than a factor of 3.  However, the mean object mass is much more sensitive to small number statistics than the median.  For example,  Calculation 1 formed 50 objects with a total mass of 5.89 M$_\odot$.  If one extra 1 M$_\odot$ star had been formed in the calculation, the mean mass would have increased by nearly 20\% whereas the median mass would have been essentially changed.  Thus, it is not clear that the mean object mass is a sensible measure of the IMF with such small numbers of objects.  That said, in the accretion/ejection IMF model, there is a tendency for the mean mass to change less than the median mass because when the turnover of the IMF moves to very low masses, the cut-off in the IMF results in fewer very low-mass objects than there would otherwise be.  However, this effect is much weaker than that observed in the hydrodynamical IMFs because the model IMFs fully populate the high-mass end of the IMF whereas the hydrodynamical calculations suffer from small number statistics at the high-mass end.

Finally, we note that although altering the minimum mass $M_{\rm min}$ by changing the metallicity of the molecular gas may not affect the form of the IMF above the cut-off directly, it may alter the IMF indirectly by changing $\overline{M}$.  Lowering the metallicity is expected to lower the density at which collapsing gas begins to heat up (hence increasing the minimum mass).  However, this is also likely to inhibit some fragmentation, lowering the number density of objects formed.  This in turn may increase $\tau_{\rm eject}$, moving the peak of the IMF to higher masses and steepening the slope of the substellar IMF.

Further calculations should be performed to test the above predictions.

\section{Conclusions}

We have presented results from the second hydrodynamical calculation to follow the collapse of a turbulent molecular cloud to form a stellar cluster while resolving fragmentation down to the opacity limit.  We compare the results with those obtained from the calculation published by Bate et al.\ (2002a,2002b,2003).  The new calculation is identical to that of Bate et al., except the progenitor cloud is nine time denser (i.e. the mean thermal Jeans mass is a factor of three lower).  

We find that the denser cloud produces a higher proportion of brown dwarfs than the original calculation.  The magnitude and sense of the dependence of the proportion of brown dwarfs on the density of the star-forming cloud reproduce the observed result that the Taurus star-forming region has a lower abundance of brown dwarfs than the Orion Trapezium Cluster (Briceno et al.\ 2002; Slesnick et al.\ 2004; Luhman 2004b).  The new calculation also produces denser groups of stars resulting in closer dynamical encounters, more severe circumstellar disc truncation, and a higher velocity dispersion than the first calculation.  

Whereas the first calculation did not produce any wide binary systems, wide binaries are produced in the new calculation when objects happen to be ejected from a small-$N$ system at roughly the time and with similar velocities and are, therefore, bound to each other.  One of the wide binaries is a binary brown dwarf system and another is a very low-mass star/brown dwarf binary.  These systems show that there may exist a population of {\it wide} ejected binary brown dwarfs.  The overall fraction of very low-mass and brown dwarf binaries produced from the two calculations is $\approx 8$\%.  This is higher than that obtained from Calculation 1 alone, but still roughly a factor of two lower than the observed fraction of binary brown dwarfs.

All objects produced by the hydrodynamical calculations begin with masses set by the opacity limit for fragmentation (approximately 0.003 M$_\odot$ in these calculations).  Those objects that end up as brown dwarfs stop accreting before they reach stellar masses because they are ejected from the dense gas soon after their formation by dynamical interactions in unstable multiple systems.  The stars are simply those objects that remain in the dense gas accreting for long enough that they exceed the hydrogen burning limit.  

Based on these calculations, we propose a simple accretion/ejection model for the origin of the IMF.  The model has three free parameters, the characteristic (median) mass which is the product of the typical protostellar accretion rate and the characteristic timescale for dynamical ejections, a dispersion in accretion rates, and the minimum brown dwarf mass which is set by the opacity limit for fragmentation.  Using values for these three parameters taken directly from the hydrodynamical calculations, the model reproduces the IMFs of the hydrodynamical calculations well.  The model predicts that the main variation of the IMF in different star-forming environments should occur in the location of the peak (in ${\rm d}N/{\rm d}\log M$) and in the substellar regime.  The peak in the IMF should occur at roughly 1/14 of the mean thermal Jeans mass in a star-forming molecular cloud.  Only a variation in the magnitude of the dispersion in the accretion rates of individual objects should alter the slope of the high-mass IMF.  A Salpeter-type slope is reproduced with an accretion rate dispersion of $\approx 0.7$ dex.  A larger dispersion results in a shallower high-mass IMF slope.

\section*{Acknowledgments}

The computations reported here were performed using the U.K.
Astrophysical Fluids Facility (UKAFF).

\end{document}